\documentclass[preprint,12pt]{elsarticle}
\usepackage[applemac]{inputenc}
\usepackage[english]{babel}
\usepackage{listings} 
\usepackage{curve2e}
\usepackage{fancyhdr}
\usepackage{lscape}
\usepackage{graphicx}
\usepackage[caption=false]{subfig}
\usepackage[fleqn]{amsmath}
\usepackage{mathtools}
\usepackage{amsfonts}
\usepackage{acronym}
\usepackage{psfrag}
\usepackage{amssymb}
\usepackage{nomencl}
\usepackage{multicol}
\usepackage{color}
\usepackage{cases}
\usepackage[standard]{frontespizio}
\usepackage{float}
\usepackage{subfloat}
\usepackage[breaklinks=true]{hyperref}
\usepackage{enumerate}
\usepackage{rotating}
\usepackage{rotfloat}
\usepackage{tabulary}
\usepackage{natbib}
\usepackage{stmaryrd}
\usepackage{setspace}
\usepackage{verbatim}
\usepackage{breakcites}
\usepackage{ar}
\usepackage{tikz,tikz-3dplot}
\usepackage[hmargin=3cm,vmargin=3.5cm]{geometry}
\usepackage{url}

\usetikzlibrary{arrows,shapes,calc}
\usetikzlibrary{decorations.pathreplacing}

\newcommand\Rey{\mbox{\textit{Re}}}  
\newcommand\Scm{\mbox{\textit{Sc}}}  
\newcommand\Pran{\mbox{\textit{Pr}}} 

\newcommand\diff{\mbox{\textit{d}}}  
\begin{document}
\journal{International Journal of Heat and Fluid Flow}
\begin{frontmatter}

\title{Direct numerical simulation of supersonic pipe flow at moderate Reynolds number}
\author{Davide Modesti$^{1,2}$ and Sergio Pirozzoli$^2$}

\address{$^1$Department of Mechanical Engineering, The University of Melbourne, Victoria 3010,Australia}
\address{$^2$Dipartimento di Ingegneria Meccanica e Aerospaziale, Sapienza Universit\`a di Roma, via Eudossiana 18, 00184 Roma, Italia}

\begin{abstract}
We study compressible turbulent flow in a circular pipe, at computationally high Reynolds number.
Classical related issues are addressed and discussed in light of the DNS data, including validity
of compressibility transformations, velocity/temperature relations, passive scalar statistics, and size of turbulent eddies.
Regarding velocity statistics, we find that Huang's transformation yields excellent universality of the
scaled Reynolds stresses distributions, whereas the transformation proposed by \citet{trettel_16}
yields better representation of the effects of strong variation of density and viscosity occurring in the
buffer layer on the mean velocity distribution. A clear logarithmic layer is recovered in terms of
transformed velocity and wall distance coordinates at the higher Reynolds number under scrutiny ($\Rey_{\tau} \approx 1000$),
whereas the core part of the flow is found to be characterized by a universal parabolic velocity profile.
Based on formal similarity between the streamwise velocity and the passive scalar transport equations,
we further propose an extension of the above compressibility transformations to also achieve universality of passive scalar statistics.
Analysis of the velocity/temperature relationship provides evidence for quadratic dependence
which is very well approximated by the thermal analogy proposed by \citet{zhang_14}.
The azimuthal velocity and scalar spectra show an organization very similar to canonical incompressible
flow, with a bump-shaped distribution across the flow scales, whose peak increases with the wall distance.
We find that the size growth effect is well accounted for through an effective length scale accounting
for the local friction velocity and for the local mean shear.
\end{abstract}

\end{frontmatter}

\section{Introduction}

Turbulent flow in circular pipes is common in many engineering applications, which cover both the
incompressible and the compressible regime.
Typical incompressible applications include transport of oil, potable, waste or irrigation water, whereas
compressible pipe flow serves as prototype for the design of air intakes used in the aircraft industry.
Several studies of incompressible pipe flow are available in the literature, which have highlighted
similarities and differences with flow in planar channels.
\citet{eggels_94} carried out experiments and direct numerical simulation (DNS) of incompressible pipe flow at 
$\Rey_b = 2 u_b R/\overline{\nu}_w=5300$ (where $u_b$ is the bulk flow velocity, $R$ is the pipe radius
and $\overline{\nu}_w$ is the kinematic viscosity at the wall, respectively),
and, in agreement with early experiments at low Reynolds number~\citep{patel_69}, found that
the mean velocity profile deviates from plane channel flow in the logarithmic and wake regions.
\citet{orlandi_97} carried out DNS of rotating and non-rotating pipe flow, and
found good agreement with the experiments and numerical simulations by~\citet{eggels_94}, in the non-rotating case. 
More recently, higher Reynolds numbers have been reached 
both in DNS~\citep{wu_12,chin_15,lee_15,ahn_15},
up to friction Reynolds number $\Rey_\tau = u_\tau R / \overline{\nu}_w \approx 3000$ 
(where $u_\tau=\sqrt{\tau_w/\overline{\rho}_w}$ is the friction velocity, $\tau_w$ the mean wall shear stress and $\overline{\rho}_w$ the mean density at the wall),
and experiments ~\citep{zagarola_98,kim_99,mckeon_04,furuichi_15}, which are typically
in the Reynolds number range $\Rey_{\tau} \sim 10^3-10^5$.
These studies generally agree that the near-wall region of turbulent pipe flow is the same
as in plane channel flow, whereas differences are found in the logarithmic and core regions.
In particular, \citet{kim_99} carried out experiments of pipe flow and observed
that the pre-multiplied streamwise velocity spectra in the logarithmic layer are characterized by a peak
at wavelengths corresponding to $24$--$28R$, and referred to these structures
as very-large-scale motions (VLSM).
\citet{wu_12} carried out DNS of incompressible pipe flow at friction Reynolds number
$\Rey_\tau \approx 700$, and found a smaller spectral peak 
than in experiments, attributing the disagreement to use of Taylor's hypothesis~\citep{del_09}. 
\citet{lee_15,ahn_15} showed that the population density
of VLSM is lower in pipe than in channel flow,
and that they survive for shorter time in pipes owing to wall confinement,
which is the likely explanation for differences of the mean velocity profile in the outer layer.
 
Compressible pipe flow has received comparatively much less attention
than the incompressible counterpart.
\citet{kjellstrom_68} carried out experiments of compressible pipe flow at 
bulk Mach number $M_b=u_b/c_w=0.1-0.3$ (based on the wall speed of sound, $c_w$).
At $M_b=0.3$ they noted compressibility 
effects on the wall shear stress, and claimed that accounting for compressibility corrections
might be necessary for accurate measurements, even at low Mach number.
\citet{sandberg_12} carried out DNS of a fully developed compressible pipe flow exiting in a co-flow,
and found good agreement of the pipe flow statistics with incompressible data up to a 
jet Mach number of $0.84$.
\citet{ghosh_06,ghosh_10} carried out the first DNS of supersonic turbulent flow in circular pipe and planar channel,
and compared the flow statistics of the two geometries at matching Mach number and friction Reynolds
number ($M_b=1.3$, $\Rey_\tau = 245$), and found 
that, as in incompressible flow, the mean flow statistics are affected by wall
curvature effects.

The effect of fluid compressibility on mean velocity and Reynolds stresses is a long-standing issue in 
wall turbulence. \citet{morkovin_62} first postulated that genuine compressibility effects 
should be negligible if density fluctuations are smaller than mean density variations ($\rho'<<\overline{\rho}$),
which became known as ``Morkovin's hypothesis''. 
This hypothesis led to several compressibility transformations for velocity
and Reynolds stresses aimed at mapping compressible flow statistics onto equivalent incompressible
ones through the mean density and viscosity profiles~\citep{vandriest_51,coleman_95,trettel_16}.
In their DNS study of pipe flow, \citet{ghosh_10} pointed out that the classical 
van Driest transformation~\citep{vandriest_51} is rather inaccurate,
whereas good universality of the density-scaled Reynolds stresses was obtained 
using the semi-local scaling introduced by \citet{huang_95}.
An extensive assessment of existing compressibility transformations
was carried out by the present authors using DNS data of plane channel flow~\citep{modesti_16}. 
The results supported effectiveness of Huang's transformation in
achieving universality of the Reynolds stress distributions.
However, better universality of the mean velocity profile, 
with respect to classical van Driest transformation~\citep{vandriest_51}, 
was obtained using 
the transformation proposed by \citet{trettel_16} (hereafter referred to as TL transformation), based on the 
enforcement invariance of the logarithmic law in the overlap layer.
The effect of compressibility on the velocity spectra, and therefore on the size of the turbulent eddies
populating the wall layer, has also been studied for long time, with 
contradictory conclusions~\citep{spina_94,smits_96,pirozzoli_11}.
Recent channel flow DNS~\citep{modesti_16} and boundary layer experiments~\citep{williams_18} 
have indeed shown that genuine compressibility effects on the typical flow scales
are small, if any, and may be conveniently accounted for through 
variable-property extension of incompressible scaling formulas.
Another subject of great interest in compressible flows is the transport of passive scalars,
an essential building block in the understanding of mixing processes and combustion.
Passive scalars in compressible wall-bounded flows have received little attention so far, mainly limited
to the case of planar channels~\citep{foysi_05}. As in incompressible flow, similarity between 
passive scalars and the streamwise velocity field is also generally taken for granted in compressible flow, 
based on similarity of the governing equations.
 
In this work we develop a DNS database for compressible flow in a circular pipe which greatly
extends the Reynolds number envelope of previous numerical studies.
Our goal is to shed light on several facets of this flow which in our opinion 
have not been adequately addressed, so far.
First, we aim at verifying the predictive power of compressibility transformations for
mean velocity and Reynolds stresses, by also comparing with available incompressible DNS data
at matching Reynolds number.
Second, we aim at establishing scaling laws for the streamwise velocity spectra in the 
outer part of the wall layer, to quantify compressibility effects on the turbulent eddies.
Third, we study the statistics and spectra of scalar fields passively advected by the fluid,
with special emphasis on identifying similarities and differences with respect to the velocity field,
and tracing possible compressibility effects.
For that purpose, we carry out DNS in the range of bulk Mach numbers $M_b=0.3$--$3$,
and of bulk Reynolds numbers $\Rey_b = 5000$--$30000$, corresponding to friction Reynolds
numbers $\Rey_\tau=180$--$1000$.

\section{Methodology}\label{sec:turb_pipe}
 
We solve the compressible Navier--Stokes equations for a perfect shock-free heat-conducting gas
in cylindrical coordinates, augmented with the transport equation for a passive scalar $\phi$,
\begin{equation}
 \frac{\partial \mathbf{w}}{\partial t}   + \frac{1}{r}\left(
 \frac{\partial \mathbf{f}_x}{\partial x} +  
 \frac{\partial \mathbf{f}_r}{\partial r} +
 \frac{1}{r}\frac{\partial \mathbf{f}_{\theta}}{\partial \theta}
 \right) - 
 \frac{\partial \mathbf{f}_x^v}{\partial x}                 -
 \frac{1}{r}\frac{\partial r\,\mathbf{f}_r^v}{\partial r}   - 
 \frac{1}{r}\frac{\partial \mathbf{f}_\theta^v}{\partial \theta} + \mathbf{S_e} - \mathbf{S_v} - \mathbf{F}= 0, 
\label{eq:N-Scyli}
\end{equation}
where $\mathbf{w}$ is the vector of the conservative variables, and $\mathbf{f}_i,\,\,
\mathbf{f}_i^v\,\, i=1,2,3$
are the convective and viscous fluxes in the coordinate directions $(x, r, \theta)$,
\begin{equation}
 \mathbf{w}=\begin{bmatrix} \rho\\ \rho u_j\\ \rho s\\\rho\phi\end{bmatrix}, \quad
 \mathbf{f}_i=\begin{bmatrix} \mathcal{R} u_i\\ \mathcal{R}\,u_i\,u_j+\mathcal{P}\,\delta_{ij} \\ \mathcal{R}\,u_i\,s\\\mathcal{R} u_i \phi\end{bmatrix},\quad
  \mathbf{f}_i^v= \begin{bmatrix}0 \\ \sigma_{ij} \\ {q_i}/{T} \\ J_i \end{bmatrix}\quad  j=1,2,3,
\label{eqn:euler_fluxes}%
\end{equation}
where $\mathcal{R} =r \rho$, $\mathcal{P} = r p$, with
$\rho$ the fluid density, $p$ the thermodynamic pressure, and $s=c_v\ln{(p\rho^{-\gamma})}$ the entropy per unit mass,
($\gamma=c_p/c_v=1.4$). 
In Eqn.~\eqref{eqn:euler_fluxes}, $\mathbf{S_e}$ and $\mathbf{S_v}$ are the Eulerian and the viscous source terms, and $\mathbf{F}$
contains the driving terms needed to keep the mass and scalar flow rate constant in time 
\begin{equation}
 \mathbf{S}_e=\frac{1}{r}\begin{bmatrix}0 \\ 0 \\ -\rho\,u_\theta^2 - p \\ \rho\,u_r\,u_{\theta} \\ 0 \\ 0\end{bmatrix},\quad 
 \mathbf{S}_v=\begin{bmatrix}0 \\ 0 \\ -{\sigma_{\theta \theta}}/{r} \\ {\sigma_{r\theta}}/{r} \\{\sigma_{\ell m}S_{\ell m}}/{T} - {q_{\ell} q_{\ell} }/{\lambda T^2} \\ 0 \end{bmatrix},\quad
 \mathbf{F}=\begin{bmatrix}0 \\ \Pi \\ 0 \\  0 \\0 \\\Phi\end{bmatrix} ,
\end{equation}
where $\sigma_{ij}=2\mu \left(S_{ij}-\frac{1}{3}\Theta\delta_{ij}\right)$,
is the viscous stress tensor, $S_{ij}$ is the strain rate tensor,
\begin{subequations}
 \begin{alignat*}{3}
 S_{xx}  &= \frac{\partial u}{\partial x}, \quad
 && S_{xr}  = \frac{1}{2}\left(\frac{\partial u_r}{\partial x}+\frac{\partial u}{\partial r}\right), \\ 
 S_{x\theta} & = \frac{1}{2} \left(\frac{\partial u_{\theta}}{\partial x}+\frac{1}{r}\frac{\partial u}{\partial \theta}\right), \quad
 && S_{rr}  = \frac{\partial u_r}{\partial r}, \\
 S_{r\theta} & = \frac{1}{2} \left(\frac{\partial u_{\theta}}{\partial r}+\frac{1}{r}\frac{\partial u_r}{\partial \theta}-\frac{u_{\theta}}{r}\right), \quad
 && S_{\theta\theta}  = \left( \frac{1}{r} \frac{\partial u_{\theta}}{\partial \theta}+\frac{u_r}{r}\right), \\
 \end{alignat*}
\end{subequations}
$\Theta$ is the dilatation,
\begin{equation}
\Theta = \frac{\partial u}{\partial x} 
+ \frac{\partial}{\partial r}\left(r\frac{\partial u_r}{\partial r}\right) 
+\frac{1}{r}\frac{\partial u_{\theta}}{\partial \theta},
\end{equation}
and $q_j$, $J_i$ are the heat flux and scalar diffusion fluxes, respectively,
\begin{subequations}
 \begin{alignat*}{3}
  &q_x  = \lambda \frac{\partial T}{\partial x},\quad
  &q_r  = \lambda \frac{\partial T}{\partial r},\quad
  &q_\theta  = \lambda \frac{1}{r}\frac{\partial T}{\partial \theta},\\
  &J_x  = \rho\alpha\frac{\partial \phi}{\partial x},\quad
  &J_r  = \rho\alpha\frac{\partial \phi}{\partial r},\quad
  &J_\theta = \rho\alpha\frac{1}{r}\frac{\partial \phi}{\partial \theta}.
 \end{alignat*}
\end{subequations}
The dependence of the viscosity coefficient on temperature is accounted for through Sutherland's law,
the thermal conductivity is defined as $\lambda =c_p\mu/\Pran$, with Prandtl number $\Pran=0.71$,
and the scalar diffusivity is expressed in nondimensional form in terms of the
Schmidt number, $\Scm = \alpha / \nu$.
Hereafter, the velocity components in coordinate directions
will be denoted as $u$, $u_r$, $u_\theta$, and the variable $y=R-r$ will be used to denote the distance from the wall.

The equations are numerically solved using a fourth-order co-located
finite-difference solver which guarantees discrete preservation of total kinetic energy from convection in the inviscid limit. For that purpose, the convective terms in Eqn.~\eqref{eq:N-Scyli}
are first expanded to a quasi-skew-symmetric form, and then discretized using standard central difference formulas~\citep{pirozzoli_10,pirozzoli_11}.
The viscous terms are expanded to Laplacian form and similarly discretized with central difference formulas.
The use of the entropy equation in place of the total energy equation
is instrumental to semi-implicit time advancement,
needed to discard the severe acoustic time step limitation in the azimuthal and wall-normal directions.
As explained in~\citet{modesti_18}, this is realized through definition of a partial flux Jacobian matrix
whereby only the effects of acoustic wave propagation are retained.
Implicit treatment of the viscous terms in
the azimuthal direction is also implemented to remove the 
viscous time step limitation at the pipe axis.
The convective time step limitation in the azimuthal direction 
must also be alleviated in practical pipe flow calculations.
For that purpose, the azimuthal
convective derivatives are evaluated at progressively 
coarsened $\theta$ resolution as the axis is approached
in such a way that the effective resolution in physical space is retained~\citep{bogey_11_jcp}.
 
The governing equations are solved in a cylindrical domain with size $L_x \times R \times 2 \pi R$.
Periodicity is imposed in the streamwise and azimuthal directions, and no-slip isothermal boundary conditions
are imposed at the wall, where we also set $\phi=0$.
Regarding the singularity at the symmetry axis, we follow \citet{mohseni_00}, and stagger 
the first collocation point at a distance $\Delta r/2$ from the axis.
Hence no special treatment of the axis is required,
provided the flow variables in the ghost points are suitably defined accounting for
symmetry/antisymmetry conditions.
The velocity and the passive scalar fields are initialized with the incompressible laminar solution,
with superposed perturbations synthesized through the digital filtering technique~\citep{klein_03},
with initially uniform density and temperature.
The flow statistics averaged in the streamwise and azimuthal directions have been collected
at equally spaced time intervals, and convergence of the flow statistics
has been checked a-posteriori.
In the following we will use
both Reynolds ($\phi=\overline{\phi} + \phi'$) and Favre ($\phi=\widetilde{\phi} + \phi''$,
$\widetilde{\phi}=\overline{\rho\phi}/\overline{\rho}$) averages, 
where the overline symbol denotes averaging in the streamwise and spanwise directions and in time.
Accordingly, the Reynolds stress components are denoted as $\tau_{ij} = \overline{\rho}\widetilde{u''_i u''_j}$. 
Quantities normalized in wall-units $u_\tau, \delta_v$ 
(where $\delta_v=\overline{\nu}_w/u_\tau$ is the viscous length scale) are denoted 
with the $+$ superscript.

\begin{figure}[]
 \centering
 \includegraphics[scale=1.1]{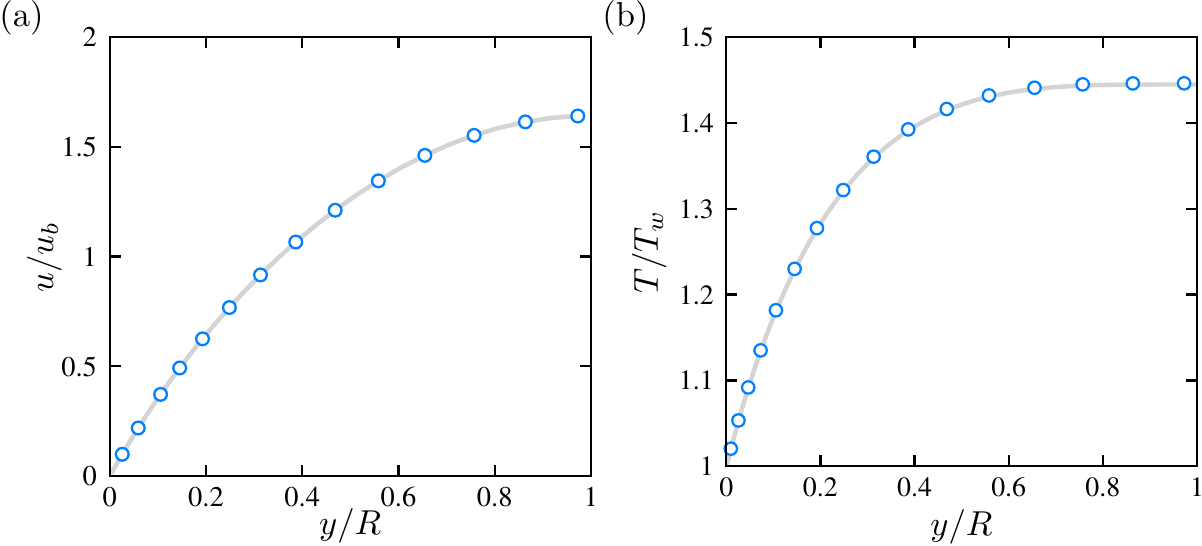}
\vskip 1em
 \caption{Laminar supersonic flow in isothermal pipe at $M_b=u_b/c_w=1.5$:
          radial profiles of velocity (a) and temperature (b).
          The DNS solution (circles) is compared with the reference one (solid lines), obtained
          using a standard ODE solver.
  }
 \label{fig:lam_pipe}
\end{figure}

The solver was preliminarily validated
for the case of supersonic laminar flow.
Figure~\ref{fig:lam_pipe} shows the velocity and temperature profiles compared
with a reference solution of the axisymmetric Navier-Stokes equations 
obtained using a standard ODE solver.
The mean velocity profile is nearly (but not exactly) parabolic, whereas
the temperature in the bulk flow is higher than at the wall temperature,
as a result of aerodynamic heating.
Overall, excellent agreement with the reference data is observed.

\begin{table}[]
\footnotesize
\begin{center}
\begin{tabular}{lcccccccccccc}
\hline
Case & $\Rey_b$ & $M_b$ & $\Rey_{\tau}$ & ${\Rey_{\tau}}_T$ & $N_x$ & $N_{r}$ & $N_{\theta}$ &${\Delta x}^+$ & $\Delta r^+_w$ & $R^+ {\Delta \theta}$ &$M_{\tau}$ & ${\Delta t}_{av}u_\tau/R$\\
\hline
  P02   &5300  & 0.2 &184 &180  & 256  & 64  &256  & 11&  4.5  &0.014 &0.0011&21.4\\
  P13   &6362  & 1.3 &235 &164  & 320  & 96  &320  & 7.3&  4.6  &0.075 &0.040&13.2\\
  P15A  &6000  & 1.5 &223 &143  & 512  & 128 &320  & 8.2&  4.4  &0.082 &0.051&17.8\\
  P15B  &14600 & 1.5 &521 & 334  & 1024 & 128 &640  & 9.6&  5.1  &0.077 &0.048&19.0\\
  P15C  &31500 & 1.5 &1030& 667 & 2048 & 256 &1280 & 9.5&  5.0  &0.070 &0.044&9.6\\
  P3    &10300 & 3.0 &524& 147 & 1024 & 128 &640 & 9.6&  5.1  &0.12 &0.15&15.7\\
\hline
\end{tabular}
\caption{Compressible pipe flow dataset.
        $M_b=u_b/c_w$ and $\Rey_b = 2 u_b R/\overline{\nu}_w$, are the bulk Mach and Reynolds number, respectively;
        $\Rey_{\tau}=R^+=R/\delta_v$ and $\Rey_{\tau T} = y_T(h)/\delta_v$ are the standard
        and transformed friction Reynolds number, as defined in Eqn.~\eqref{eq:RetauT}.
$N_x$, $N_r$, $N_{\theta}$ are the number of mesh points in the streamwise, radial
and azimuthal direction, and $\Delta x^+$, $\Delta r^+_w$, $R^+ \Delta \theta$ are the respective inner-scaled mesh spacings at the wall.
$M_\tau=u_\tau/c_w$ is the friction Mach number.
The computational domain length is $L_x=6\pi R$ for all cases, with the exception of flow case P13 having 
$L_x=10 R$, for consistency with~\citet{ghosh_10}.
$\Delta t_{av}$ is the time averaging interval.}
\label{tab:testcases_pipe}%
\end{center}
\end{table}%

\begin{figure}[]
 \centering
 \includegraphics[scale=1.1]{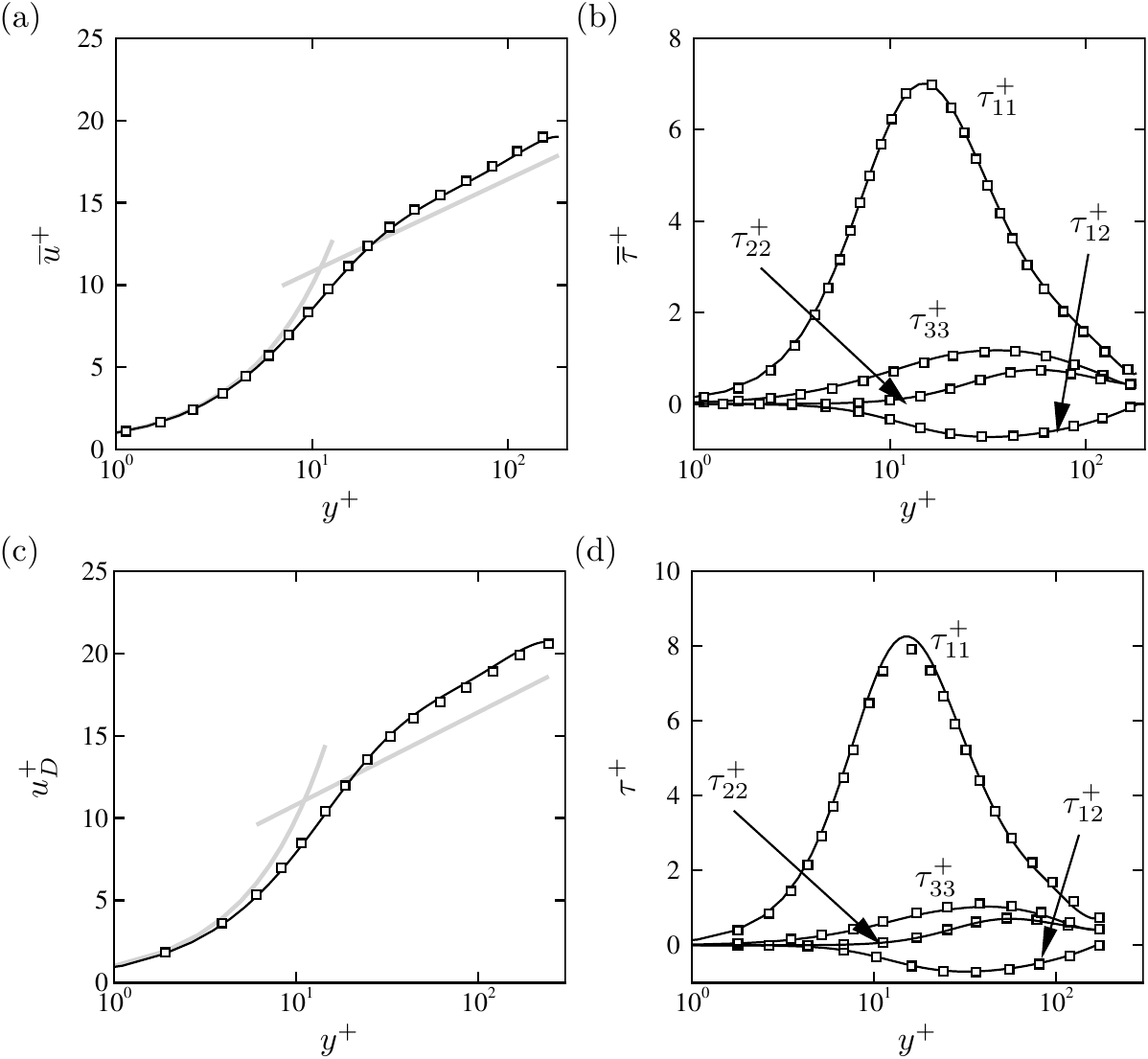}
 \vskip 1em
 \caption{Comparison of mean velocity profiles (left column) and turbulent stresses (right column) with reference literature data.
    (a), (b) comparison of P02 flow case with incompressible pipe DNS data by \citep{wu_08};
    (c), (d) comparison of P13 flow case with compressible pipe DNS data by \citep{ghosh_10},
    where $u_D$ is the van Driest transformed velocity profile, defined in Eqn.~\eqref{eq:uvd}.
    Solid lines denote the present DNS results, and symbols the reference data. 
    The thick gray lines in panels (a), (c) indicate the standard wall law,
    $\overline{u}^+=y^+$, $\overline{u}^+=1/k\log{(y^+)}+B$, $k=0.41$, $B=5.2$. 
  }
 \label{fig:valid}
\end{figure}

A list of the computed flow cases is reported in Tab.~\ref{tab:testcases_pipe}.
Three DNS have been carried out at bulk Mach number $M_b=1.5$ by changing the Reynolds number up to $\Rey_\tau \approx 1000$, and one DNS has been carried out at $M_b=3$, $\Rey_\tau=500$.
Two additional DNS have been carried out, one at subsonic conditions to match reference incompressible DNS data by~\citet{wu_08}, 
and one at $M_b=1.3$, to match the conditions considered by~\citet{ghosh_10}.
Figure~\ref{fig:valid} shows a comparison of the mean velocity profile and of the Reynolds stresses distributions.
Agreement of the P02 flow case with reference data is excellent, and no compressibility effects 
are found at this low Mach number, as expected.
Agreement is also good in general terms with the compressible pipe DNS by \citet{ghosh_10}.
Significant offset with respect to the standard wall law is observed in this case, 
mainly because of poor effectiveness of van Driest transformation
in accounting for density and viscosity variations in the near-wall layer, as discussed later on.
Some differences with respect to the reference data are found as regards the peak of $\tau_{11}$.
This is likely an effect of grid resolution, as \citet{ghosh_10} used 
$\Delta x^+=9.5$, $R^+ \Delta \theta \approx 12$, jointly with upwind-biased discretization.
Inaccurate prediction of the wall shear stress is also the likely cause for the slight shift 
between the two velocity profiles in panel (c). Indeed, for the same bulk Reynolds number 
we find $\Rey_{\tau}=235$, whereas \citet{ghosh_10} report $\Rey_{\tau}=245$.
Hence, lower logarithmic intercept is expected in our case.

\section{Results}

\subsection{Velocity statistics}

Several attempts have been made in the past to remove compressibility
effects from statistics of wall-bounded flows, starting from
the classical analysis of \citet{vandriest_51}.
Mean momentum balance in turbulent pipe flow requires
\begin{equation}
  \overline{\mu}\frac{\mathrm{d} \widetilde{u}}{\mathrm{d} y} 
 - \bar{\rho}\widetilde{u''v''} =
 \overline{\rho}_w u_{\tau}^2 \left(1- \eta\right),
\label{eq:mmb_pipe}
\end{equation}
where $\eta=y/R$ is the outer-scaled wall-normal coordinate.
Away from the wall, molecular viscosity is negligible, and
further assuming $\eta << 1$, constancy of the turbulent stress follows, hence
\begin{equation}
-{\widetilde{u'' v''}}
\approx \left( \frac {\overline{\rho}_w}{\overline{\rho}} \right) \,
u^2_{\tau} , \label{eq:reysc}
\end{equation}
which shows that `compressible' stresses should be scaled by the local mean density to recover
the incompressible behavior, hence universality of the transformed stress tensor is expected
\begin{equation}
{\tau_{ij}}_D = \frac{\overline{\rho}}{\overline{\rho}_w} \, {\tau_{ij}}. \label{eq:taud}
\end{equation}
Mixing length modeling of the turbulent shear stress further leads to the classical overlap-layer equation
\begin{equation}
\frac{\diff u_D}{\diff y} = \frac {u_{\tau}}{k y}, \label{eq:loglaw}
\end{equation}
where $k$ is the von K\'arm\'an constant and $u_D$
the van Driest transformed velocity, defined as
\begin{equation}
 {u_{D}} = \int_0^{\bar{u}}
 {\left(\frac{\bar{\rho}}{\bar{\rho}_w}\right)}^{1/2}
 \mathrm{d}\widetilde{u}.
\label{eq:uvd}
\end{equation}
Integration of \eqref{eq:loglaw} directly yields a logarithmic layer for the
transformed velocity field with the same slope as in the incompressible case,
however with an additive constant which may in general vary with both Reynolds
and Mach number.
It should be noted that Eqn.~\eqref{eq:loglaw} is obtained by neglecting viscous
transport effects and mean viscosity variations which are dominant in the viscous sublayer,
hence the van Driest transformation 
cannot be expected to yield universality in the entire wall layer. 
Since the error is mainly concentrated in the near-wall region,
it may be expected that the van Driest transformation recovers its accuracy at sufficiently high Reynolds number,
as the thickness of the viscous sublayer becomes negligible with respect to the wall layer thickness.
It is also clear that in the case of adiabatic walls, since
$\overline{\rho}/\overline{\rho}_w\approx 1$ in the near-wall region,
van Driest transformation yields accurate results throughout the wall
layer~\citep{pirozzoli_11}.

Failure of van Driest transformation was highlighted in previous
studies, in which alternative transformations were proposed to map the whole wall layer.
Empirical evidence~\citep{huang_95} suggested scaling the wall-normal coordinate
based on semi-local wall units, defined as
\begin{equation}
u^*_{\tau} = \sqrt{\tau_w/\overline{\rho}}, \quad \delta_v^* = \overline{\nu}/u^*_{\tau}, \label{eq:SL}
\end{equation}
yields better collapse of the flow statistics across the Mach number range.
As it can be readily shown, this is equivalent to mapping the wall-normal coordinate
according to
\begin{equation}
y_H = \left( \frac{\rho_w}{\overline{\rho}} \right)^{1/2} \frac{\overline{\nu}_w}{\overline{\nu}}y .  \label{eq:yH}
\end{equation}
\citet{trettel_16} further investigated
failure of the van Driest transformation in the case of cold walls,
showing that Huang's semi-local
scaling is actually rooted in arguments of mean momentum balance,
and derived a novel velocity transformation which by construction satisfies universality
of the turbulent stresses.
The TL transformation can be expressed in terms of stretched space and velocity coordinates, defined as
\begin{equation}
y_T(y)=\int_0^y f_T(\eta) \, {\mathrm d}\eta, \quad
u_T(y)=\int_0^y g_T(\eta) \, \frac{\diff \tilde{u}}{\diff \eta} {\mathrm d}\eta,
\label{eq:transfrules}
\end{equation}
where
\begin{equation}
f_T(y)=\frac {\diff}{\diff y} \left( \frac y{P^{1/2} N} \right),\quad
g_T(y)=P N \frac {\diff}{\diff y} \left( \frac y{P^{1/2} N} \right),
\label{eq:kernels}
\end{equation}
with $N(y)=\overline{\nu}/\overline{\nu}_w$, $P(y)=\overline{\rho}/\overline{\rho}_w$.
It is important to note that the viscous-scaled wall normal coordinate
defined in the first of Eqn.~\eqref{eq:transfrules} coincides with 
Huang's semi-local scaling given in Eqn.~\eqref{eq:yH}.
 
\begin{figure}[]
 \centering
 \includegraphics[scale=1.1]{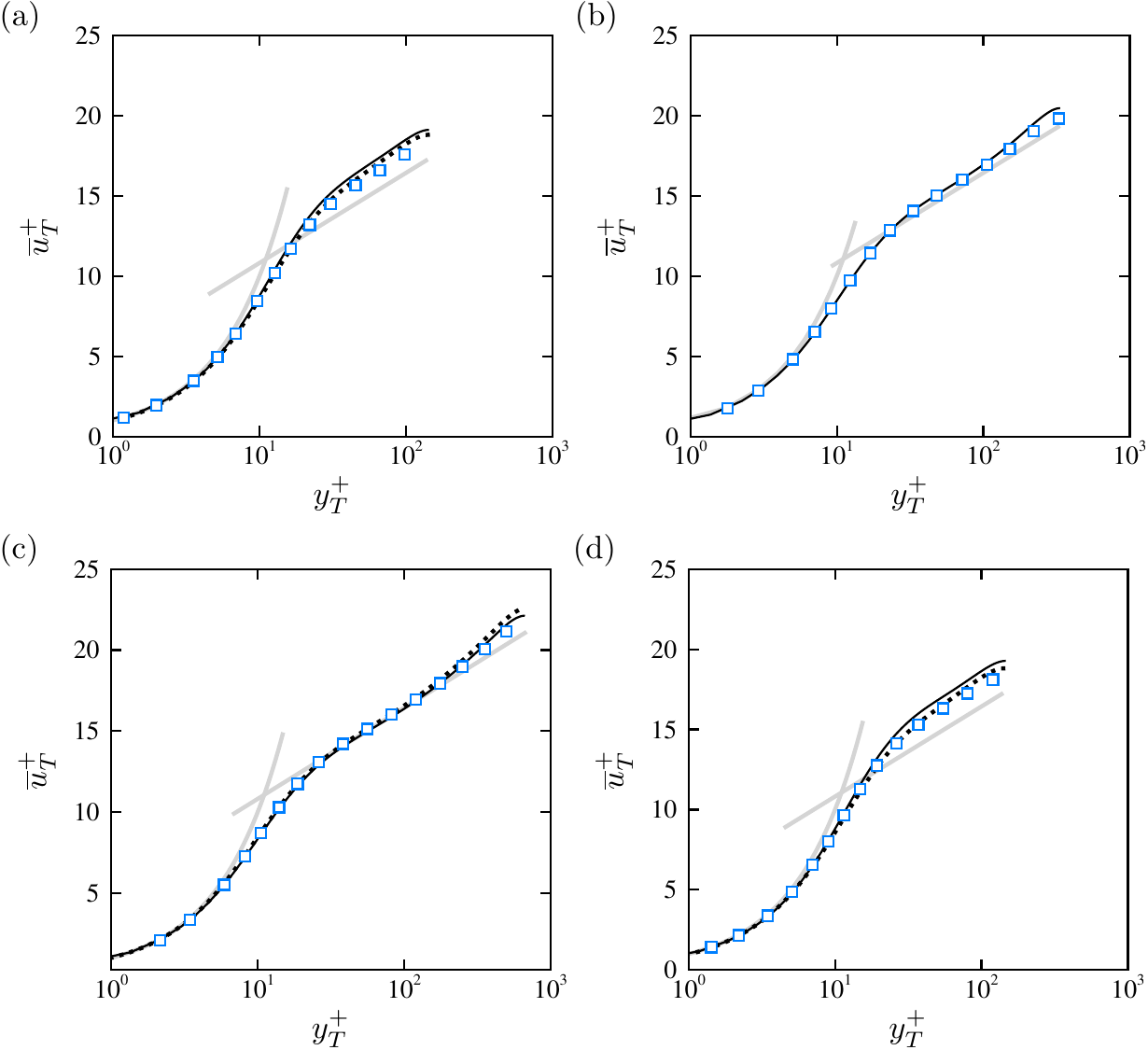}
 \vskip1.em
 \caption{Mean velocity profiles 
          for flow cases P15A (a), P15B (b), P15C (c), P3 (d), upon TL transformation. 
          The present DNS data (solid lines)
          are compared with compressible channel data \citep[squares]{modesti_16} at 
          ${\Rey_{\tau}}_T = 141$ (a)-(d), $333$ (b), $677$ (c), and
          incompressible pipe data (dotted lines) at $\Rey_{\tau}=140$ \citep[(a), (d)]{verzicco_96},
          and at $\Rey_{\tau}=680$ \citep[(c)]{wu_08}.
          The thick gray lines indicate the standard wall,
          $u^+=y^+$, $u^+=1/0.41\log{(y^+)}+5.2$.
  }
 \label{fig:vel_trettel_pipe}
\end{figure}

Figure~\ref{fig:vel_trettel_pipe} shows the mean transformed velocity profiles 
obtained from DNS upon TL transformation, compared with reference compressible channel flow data~\citep{modesti_16},
and incompressible pipe flow data~\citep{verzicco_96,wu_08}.
For the sake of comparison, data are taken at similar values of the
friction Reynolds number based on the transformed outer length scale, namely
\begin{equation}
{\Rey_{\tau}}_T = y_T(R)/\delta_v. \label{eq:RetauT}
\end{equation}
Figure~\ref{fig:vel_trettel_pipe} shows general success of the TL transformation in 
mapping compressible DNS results back to incompressible ones across the Reynolds number range,
with minor deviations at low $\Rey$. This is not the case of the van Driest transformation
(the results are not shown, but one may refer to figure~\ref{fig:valid}(c)), yielding large deviations in the buffer layer which persist into the outer layer.
As in incompressible flows, large upward excursions from the standard log-law are found at low
Reynolds number, whereas near logarithmic behavior is recovered starting from the P15C simulation.
Comparison with channel flow data shows very similar distributions in the inner layer, 
with a slightly stronger wake region, again as in incompressible flow.

\begin{figure}[]
 \includegraphics[scale=1.1]{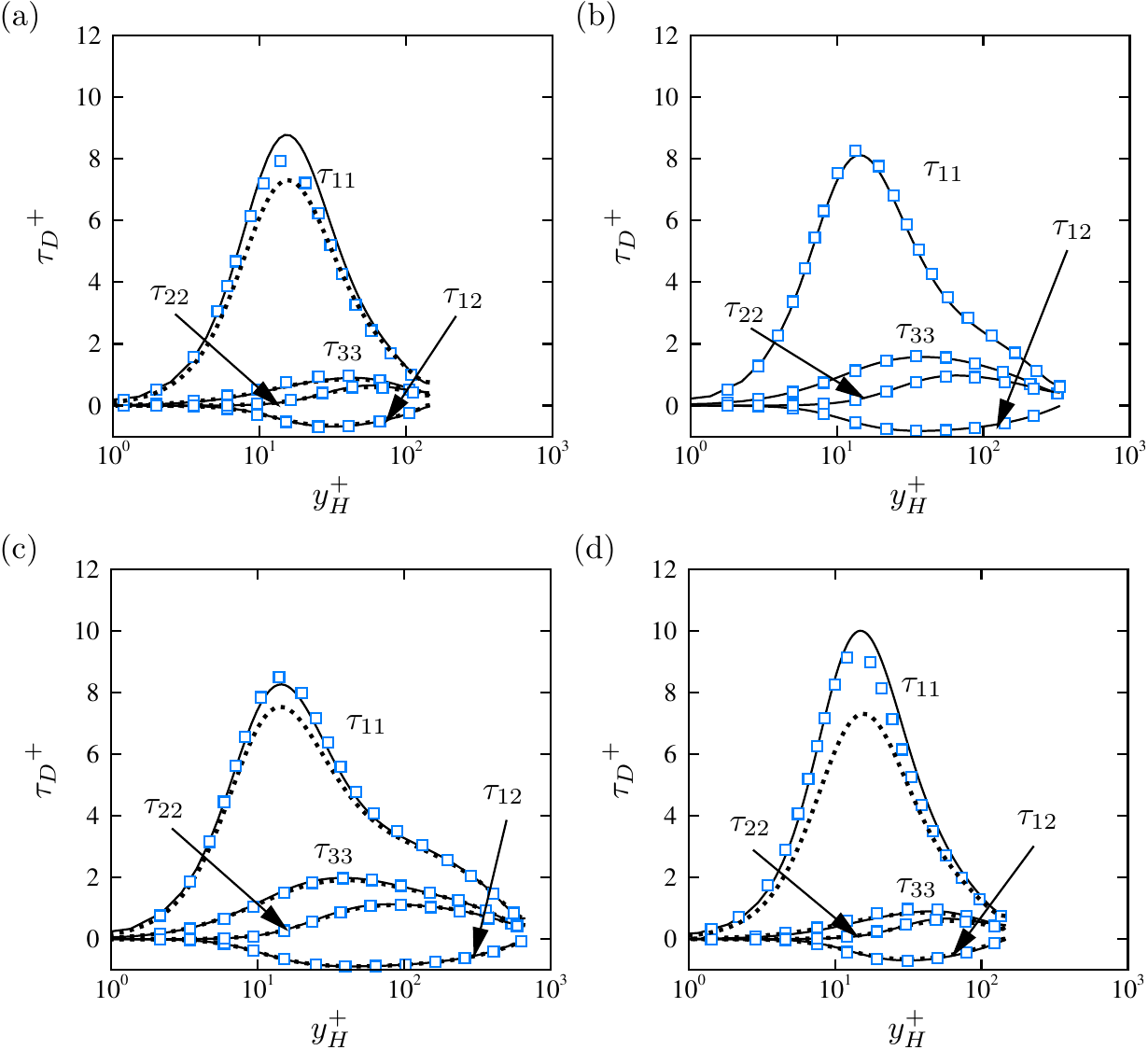}
\vskip 1em
 \caption{Distributions of transformed turbulent stresses
          for flow cases P15A (a), P15B (b), P15C (c), P3 (d).
          The present DNS data (solid lines)
          are compared with compressible channel data \citep[squares]{modesti_16} at 
          ${\Rey_{\tau}}_T = 141$ (a)-(d), $333$ (b), $677$ (c), and
          incompressible pipe data (dotted lines) at $\Rey_{\tau}=140$ \citep[(a), (d)]{verzicco_96},
          and at $\Rey_{\tau}=680$ \citep[(c)]{wu_08}.
  }
 \label{fig:rms_trettel_pipe}
\end{figure}

Similar conclusions can be drawn from scrutiny of the Reynolds stresses transformed 
according to Eqn.~\eqref{eq:taud}, and shown in Fig.~\ref{fig:rms_trettel_pipe}. 
All the Reynolds stress components actually collapse on the incompressible pipe
distributions, with the exception of the buffer peak of $\tau_{11}$,
whose offset however decreases with the Reynolds number.
The peak overshoot of $\tau_{11}$ in compressible flow calculations
is a well known feature also in boundary layers~\citep{pirozzoli_13} and channels~\citep{modesti_16},
and it is likely due to genuine compressibility effects, not fully explained so far.
Close similarity with the stresses distributions in compressible channel is also observed,
with differences again limited to the peak of $\tau_{11}$, and vanishing at higher Reynolds number.
This is consistent with observations made regarding Reynolds number effects 
in the incompressible regime~\citep{chin_14}.

\begin{figure}[]
 \centering
 \includegraphics[scale=1.1]{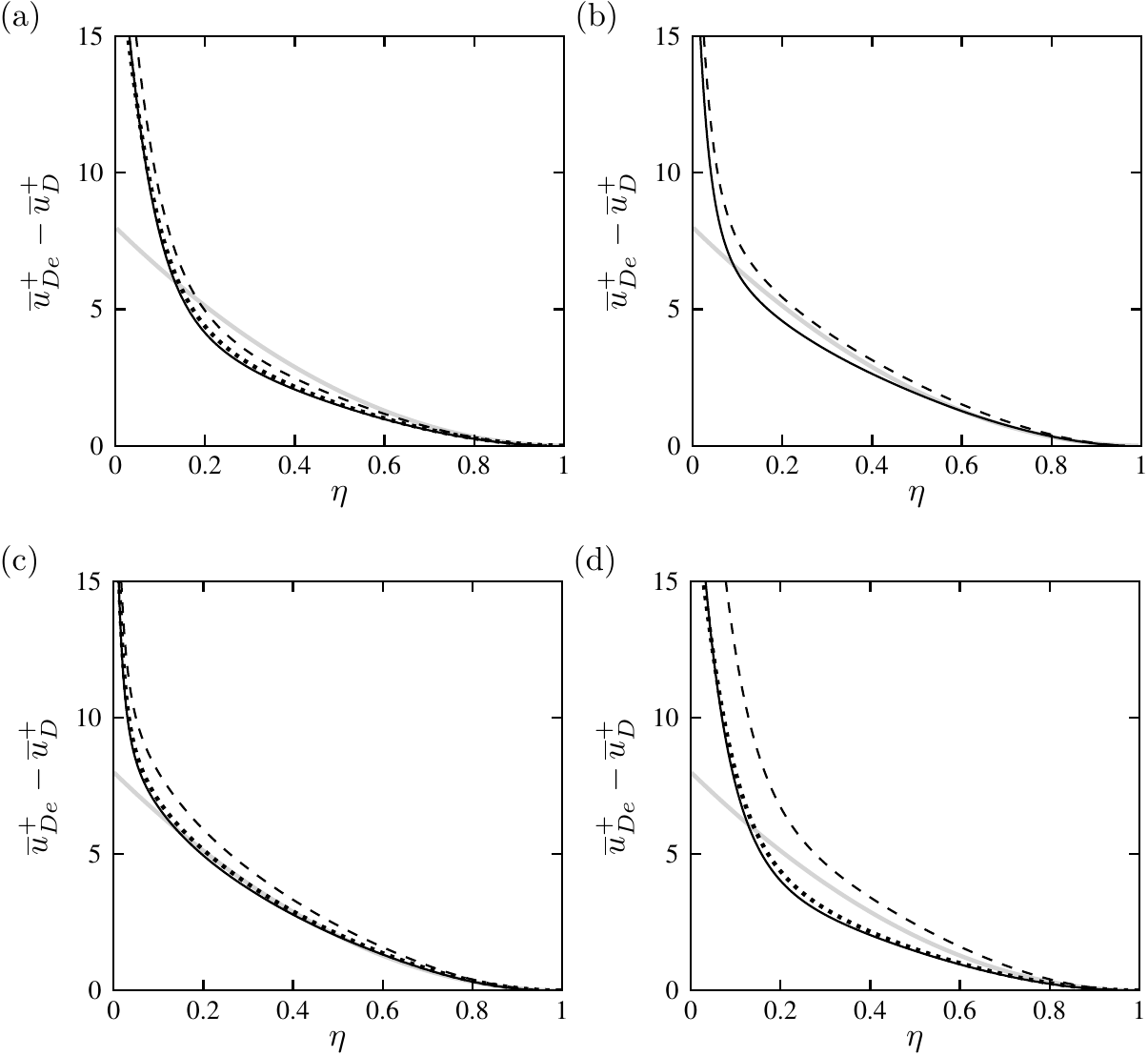}
 \vskip 1em
 \caption{Mean streamwise velocity profiles in defect form, as a function
           of $\eta=y/R$ for flow cases
           P15A (a), P15B (b), P15C (c) and P3 (d).
           Untransformed velocity profile (dashed lines),
           van Driest velocity profile (solid lines) and
           incompressible pipe data (dotted lines) at $\Rey_{\tau}=140$ \citep[(a), (d)]{verzicco_96},
           and at $\Rey_{\tau}=680$ \citep[(c)]{wu_08}.
           The gray lines denote the parabolic velocity profile 
           given in Eqn.~\eqref{eq:outpara}, with $c_\mu=0.0625$.
           The subscript 'e' denotes properties at the pipe centerline.            
  }
 \label{fig:vel_out_pipe}
\end{figure}

Closer scrutiny of the core part of the flow can be gained
by inspecting the mean velocity profiles in defect form, as given in Fig.~\ref{fig:vel_out_pipe}.
A parabolic law for the core velocity profile of incompressible pipe flow
was derived by \citet{pirozzoli_14}.
The derivation stems from the idea that the outer-layer
turbulent eddies are not directly affected by the presence of the wall, and their
size should hence scale with the pipe radius and with the typical
eddy velocity scale (namely the friction velocity),
hence it follows that the relevant eddy viscosity is
\begin{equation}
\nu_t = c_{\mu}u_{\tau}R,
\end{equation}
where $c_{\mu}$ is a suitable constant.
This argument may be extended to compressible flow based on the assumption
that in the presence of mean density variations the effective velocity
scale is $u_{\tau}^*$ (as defined in Eqn.~\eqref{eq:SL}) rather than $u_{\tau}$, which yields
the eddy viscosity
\begin{equation}
 \nu_t = c_{\mu} u_{\tau}^* R . \label{eq:nutc}
\end{equation}
From Eqn.~\eqref{eq:mmb_pipe}, neglecting the viscous term and using
the eddy viscosity~\eqref{eq:nutc}, one readily obtains
\begin{equation}
 \frac{\mathrm{d} \widetilde{u}^+}{\mathrm{d}\eta} = 
 \frac{1}{c_{\mu}}{\left(\frac{\overline{\rho}_w}{\bar{\rho}}\right)}^{1/2}
 \left(1-\eta\right) ,
\label{eq:outer_parabola}
\end{equation}
from which it follows that the van-Driest-transformed velocity
should follow a universal parabolic law in the core part
of the pipe 
\begin{equation}
 {u_{D}^+} -{u_{D}^+}_e = -\frac{1}{2c_{\mu}}\left(1-\eta\right)^2,  
 \label{eq:outpara}
\end{equation}
where ${u}_{De}$ is the transformed centerline velocity.
Outer defect profiles obtained with van Driest transformation are given
in Fig.~\ref{fig:vel_out_pipe}. 
Comparison with incompressible DNS (dotted lines) shows very good agreement
throughout the outer layer, irrespective of the Reynolds and Mach number.
The DNS data are consistent with the prediction of Eqn.~\eqref{eq:outpara}
around the pipe centerline,
the range of validity of the parabolic fit extending to more than half pipe radius
at sufficiently high Reynolds number.
No evident compressibility effects are observed on the parabolic law constant,
which in fact coincides with its incompressible value, namely $c_{\mu} \approx 0.0625$~\citep{pirozzoli_14}.
Similar results would be obtained with the TL transformation, as
density and viscosity variations in the outer wall layer are but moderate.

\subsection{Temperature/velocity relationship}
 
\begin{figure}[]
 \centering
 \includegraphics[scale=1.1]{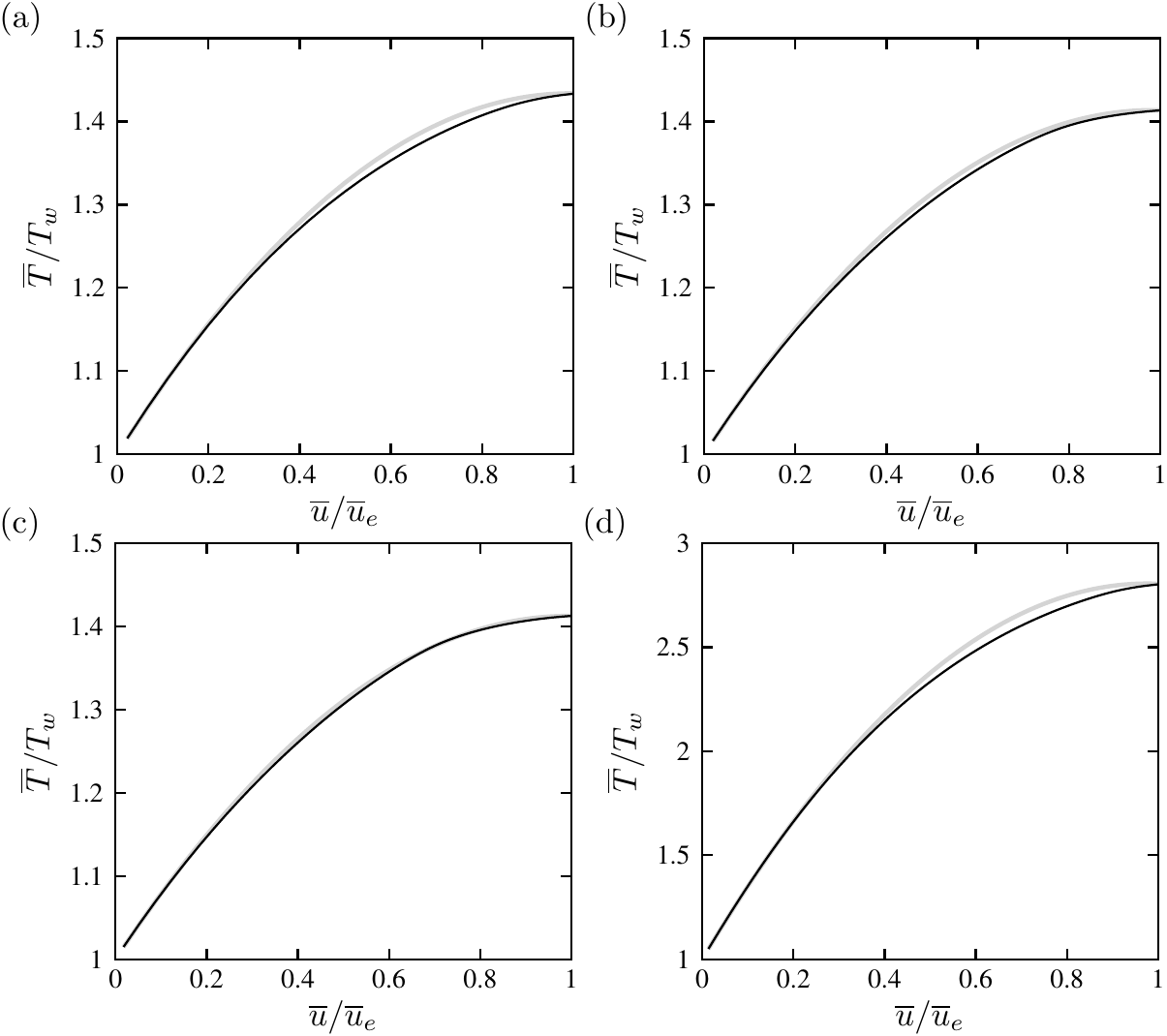}
 \vskip1.em
 \caption{Mean temperature as a function of mean streamwise velocity
          for flow cases P15A (a),P15B (b), P15C (c), P3 (d). 
          Solid lines denote DNS data, and the thick grey lines the predictions of Eqn.~\eqref{eq:zhang}. 
  }
 \label{fig:temp_pipe}
\end{figure}

An important subject in the study of compressible wall-bounded turbulence
is that of temperature/velocity relations. Successful definition
of a relation between temperature and velocity in fact allows
to use inverse compressibility transformations 
(i.e. mapping incompressible velocity distributions to compressible ones),
to derive explicit formulas for the 
friction coefficient~\citep{smits_96}. The classical temperature/velocity relation by~\citet{walz_59}
has proven its accuracy in the case of adiabatic walls~\citep{duan_10},
but it is found to fail in the case of isothermal walls~\citep{modesti_16}.
Recently, \citet{zhang_14} derived the following generalized temperature/velocity relation,
\begin{equation}
\frac{\widetilde{T}}{T_w} = 1 + \frac{T_{rg}-T_w}{T_w}\frac{\widetilde{u}}{\widetilde{u}_e} 
 + \frac{\widetilde{T}_e-T_{rg}}{T_w}\left(\frac{\widetilde{u}}{\widetilde{u}_e}\right)^2,
\label{eq:zhang}
\end{equation}
where $T_{rg}=\tilde{T}_e + r_g \tilde{u}_e^2/(2c_p)$ is a generalized recovery temperature,
$r_g = 2 c_p (T_w-\tilde{T}_e) / \tilde{u}_e^2 - 2 \Pran q_w / (\tilde{u}_e \tau_w)$ is a generalized recovery factor,
and $u_e$ and $T_e$ are the external values of velocity and temperature,
here interpreted as the pipe centerline values.
Equation~\eqref{eq:zhang} explicitly takes into account
the wall heat flux $q_w$, and it reduces to the Walz relation
in the case of adiabatic walls.
Figure~\ref{fig:temp_pipe} shows mean temperature as a function of mean
velocity for all points along the pipe radial direction.
Nearly perfect superposition of the supersonic DNS data with
the predictions of equation~\eqref{eq:zhang} is observed, 
especially at the highest $\Rey$.
Integration of the inverse compressibility transformations~\eqref{eq:transfrules} 
together with relationship \eqref{eq:zhang} can then be exploited to
reconstruct the mean velocity profile for assigned values of the bulk Reynolds and 
Mach number.

\subsection{Passive scalars}

\begin{figure}[]
 \centering
 \includegraphics[scale=1.1]{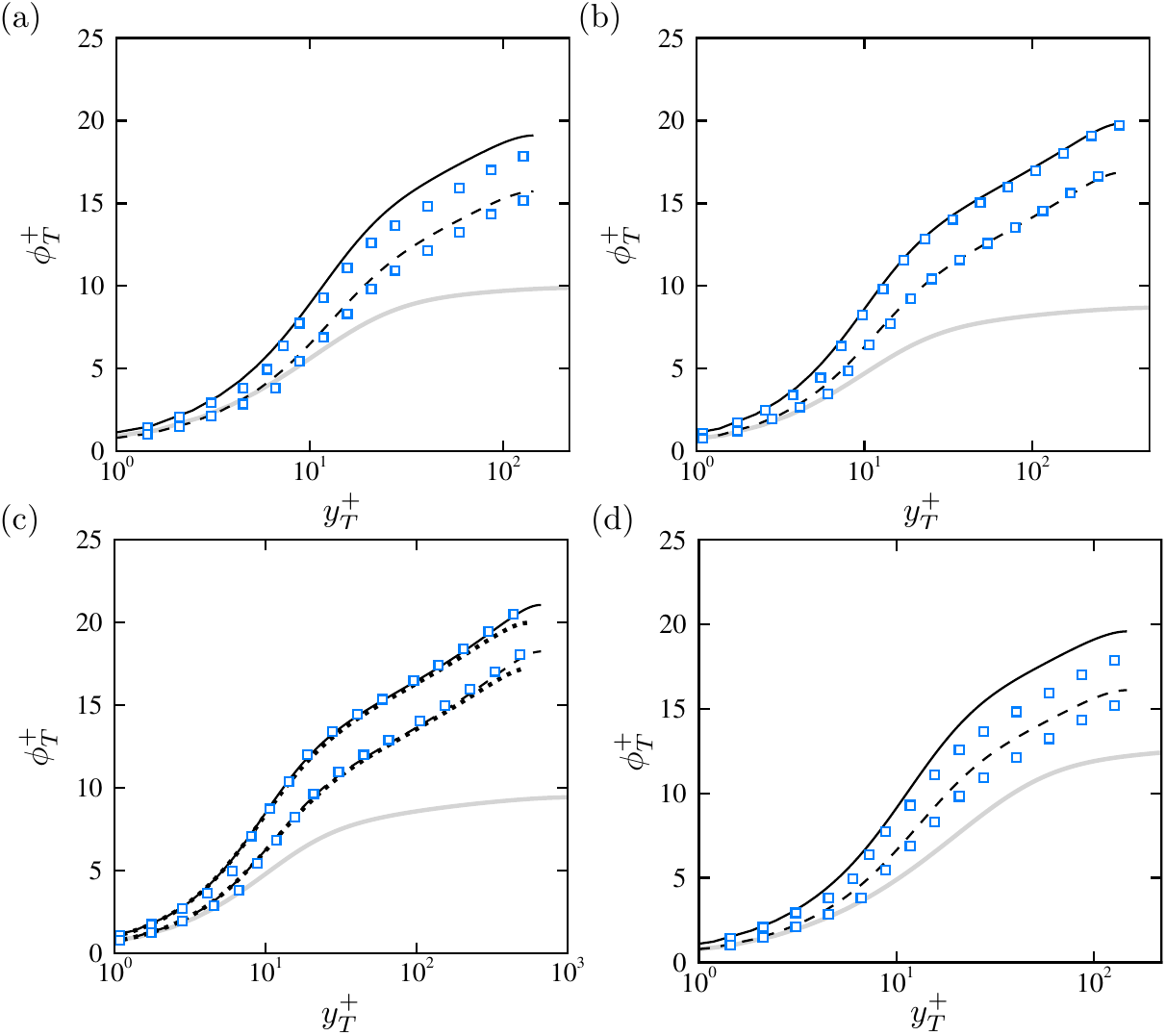}
 \vskip1.em
 \caption{Inner-scaled passive scalar mean profiles transformed according to Eqn.~\eqref{eq:scalar_trett},
          at $\Scm=1$ (solid lines) and $\Scm=0.71$ (dashed lines),
          for flow cases P15A (a), P15B (b), P15C (c), P3 (d).
          Experimental data fit for incompressible flow~\citet{kader_81} at corresponding  
          values of ${\Rey_{\tau}}_T$ are shown with blue squares.
          Dotted lines in panel (c) indicate DNS data in incompressible channel flow~\citep{pirozzoli_16}
          at $\Rey_{\tau}=548$. The thick gray lines denote inner-scaled 
          temperature profiles, $T^+=(T-T_w)/T_\tau$. 
  }
 \label{fig:sca_pipe_trettel}
\end{figure}

The statistics of passive scalar fields at Schmidt number $\Scm=0.71, 1$,
are herein presented, with the main goal of comparing with 
the behavior of the streamwise velocity and temperature fields.
The mean passive scalar balance equation reads 
\begin{equation}
  \overline{\rho}\overline{\alpha}\frac{\mathrm{d} \widetilde{\phi}}{\mathrm{d} y} 
 - \bar{\rho}\widetilde{\phi''v''} =
 \overline{\rho}_w \phi_{\tau}^2 \left(1-\eta\right),
\label{eq:mpassive_pipe}
\end{equation}
where $\phi_{\tau}=\overline{\alpha}_w/{u_\tau}(\partial \widetilde{\phi}/\partial y)_w$ 
is the reference friction value. It is clear that for $\Scm=1$, Eqn.~\eqref{eq:mpassive_pipe}
is identical to the mean momentum balance equation, Eqn.~\eqref{eq:mmb_pipe}, upon
replacement of $\phi$ with $u$.
This observation leads to expect that the TL transformation
for velocity also applies to Eqn.~\eqref{eq:mpassive_pipe}, thus allowing to map
passive scalar distributions to their incompressible counterparts.
Hence, by analogy with Eqn.~\eqref{eq:transfrules} we propose a transformation for $\phi$
\begin{equation}
\phi_T(y)=\int_0^y g_T(\eta) \, \frac{\diff \tilde{\phi}}{\diff \eta} {\mathrm d}\eta,
\label{eq:scalar_trett}
\end{equation} 
where $g_T$ is the same mapping function used for $u$, as defined in Eqn.~\eqref{eq:kernels}.
Figure~\ref{fig:sca_pipe_trettel} shows the inner-scaled transformed passive scalar profiles,
compared with the correlations developed by~\citet{kader_81} for incompressible pipe flow,
which include a logarithmic layer
with von K\'arm\'an constant $k_{\phi} = 0.47$, and an additive constant varying with the Schmidt number.
Very good agreement is indeed recovered at the higher Reynolds number,
which is not surprising as Kader's correlation was based on data fitting of
high-Reynolds-number experiments. Hence, good performance of the TL transformation is 
also confirmed for passive scalar fields. Comparison with the mean velocity profiles 
given in Fig.~\ref{fig:vel_trettel_pipe} shows similar behavior of mean velocity and mean passive
scalar at $\Scm=1$, with the small but important difference that the log-law constant is 
$k \approx 0.41$ for $u$, and $k \approx 0.47$ for $\phi$.
Figure~\ref{fig:sca_pipe_trettel}(c) also reports passive scalar distributions in incompressible channel
flow at matching $\Rey_{\tau T}$~\citep{pirozzoli_16}. 
As for the velocity field, the agreement is excellent, with the exception of the wake region,
where pipe flow exhibits a stronger wake component.

Temperature profiles in wall units are also shown in Fig.~\ref{fig:sca_pipe_trettel}
to highlight differences with passive scalar profiles. It should be noted that temperature 
here is scaled with respect to its reference friction value, $T_\tau =\lambda_w/(\rho_w c_p u_\tau) ({\partial \widetilde{T}}/{\partial y})_w$, where $\lambda_w$ is the fluid thermal conductivity at the wall.
As expected, the inner-scaled temperature profiles well agree with the passive scalar profiles
corresponding to $\Scm=0.71$ in the viscous sublayer, but they exhibit strong deviations
thereof further away from the wall. Reasons for this difference may be explained by 
comparing the mean passive scalar and the mean temperature equations.
The current flow case, with isothermal walls, shows perfect similarity of the equations 
with the exception 
of the forcing term, which is by construction spatially uniform in the equation for $\phi$,
whereas the viscous dissipation term in the temperature equation is
mainly concentrated in the vicinity of the wall.

\begin{figure}[]
 \centering
 \includegraphics[scale=1.1]{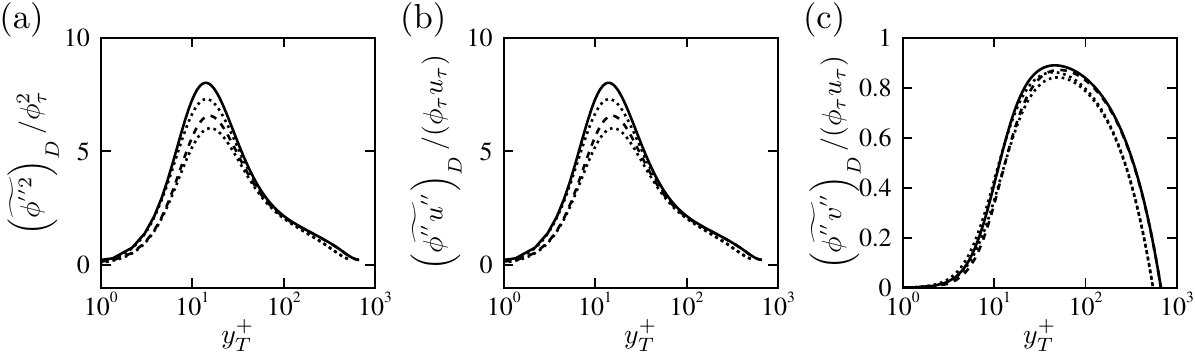}
 \caption{
          Passive scalars fluctuations and velocity/scalar correlations 
          transformed according to Eqn.~\eqref{eq:scalar_fluct}, 
          for case P15C, at $\Scm=1$ (solid lines) and $\Scm=0.71$ (dashed lines).
          The dotted lines denote the corresponding values in incompressible channel flow,
          at $\Rey_{\tau}=548$~\citep{pirozzoli_16}.
  }
 \label{fig:scal_fluc_pipe}
\end{figure}

Analogy of the mean momentum balance equation and the mean passive scalar equation
further suggests that van Driest transformation may be used to map passive scalar variances
and velocity/scalar correlations.
From similarity with Eqn.~\eqref{eq:taud}, we then consider the following mappings
\begin{equation}
 \left(\widetilde{\phi^{''2}}\right)_D = \frac{\overline{\rho}}{\overline{\rho}_w}\widetilde{\phi^{''2}}, \quad
 \left( \widetilde{\phi^{''}u^{''}}\right)_D = \frac{\overline{\rho}}{\overline{\rho}_w} \left( \widetilde{\phi^{''}u^{''}} \right), \quad
 \left( \widetilde{\phi^{''}v^{''}} \right)_D = \frac{\overline{\rho}}{\overline{\rho}_w} \left( \widetilde{\phi^{''}v^{''}} \right) .
\label{eq:scalar_fluct}
\end{equation} 
Figure~\ref{fig:scal_fluc_pipe} shows passive scalar fluctuations and scalar/velocity correlations for flow case P15C, transformed as in Eqn.~\eqref{eq:scalar_fluct}. The statistics are compared with 
passive scalar statistics in incompressible channel flow~\citep{pirozzoli_16} at approximately matching $\Rey_{\tau T}$.
Very good coincidence with the incompressible distributions is observed for the
vertical scalar flux, see panel (c), whereas differences are observed for the near-peak of the
scalar variance and of the streamwise scalar flux, which are consistently higher in the
scaled compressible data. Again, this in similarity with the observations made regarding the 
buffer layer peak of the streamwise velocity variance when commenting Fig.~\ref{fig:rms_trettel_pipe}.
This is also an important clue that pressure, which is absent in passive scalar transport,
does not play an important role in the breakdown of density scaling for the streamwise velocity variance
peak, and reasons should be traced elsewhere.

\begin{figure}[]
 \centering
 \includegraphics[scale=1.1]{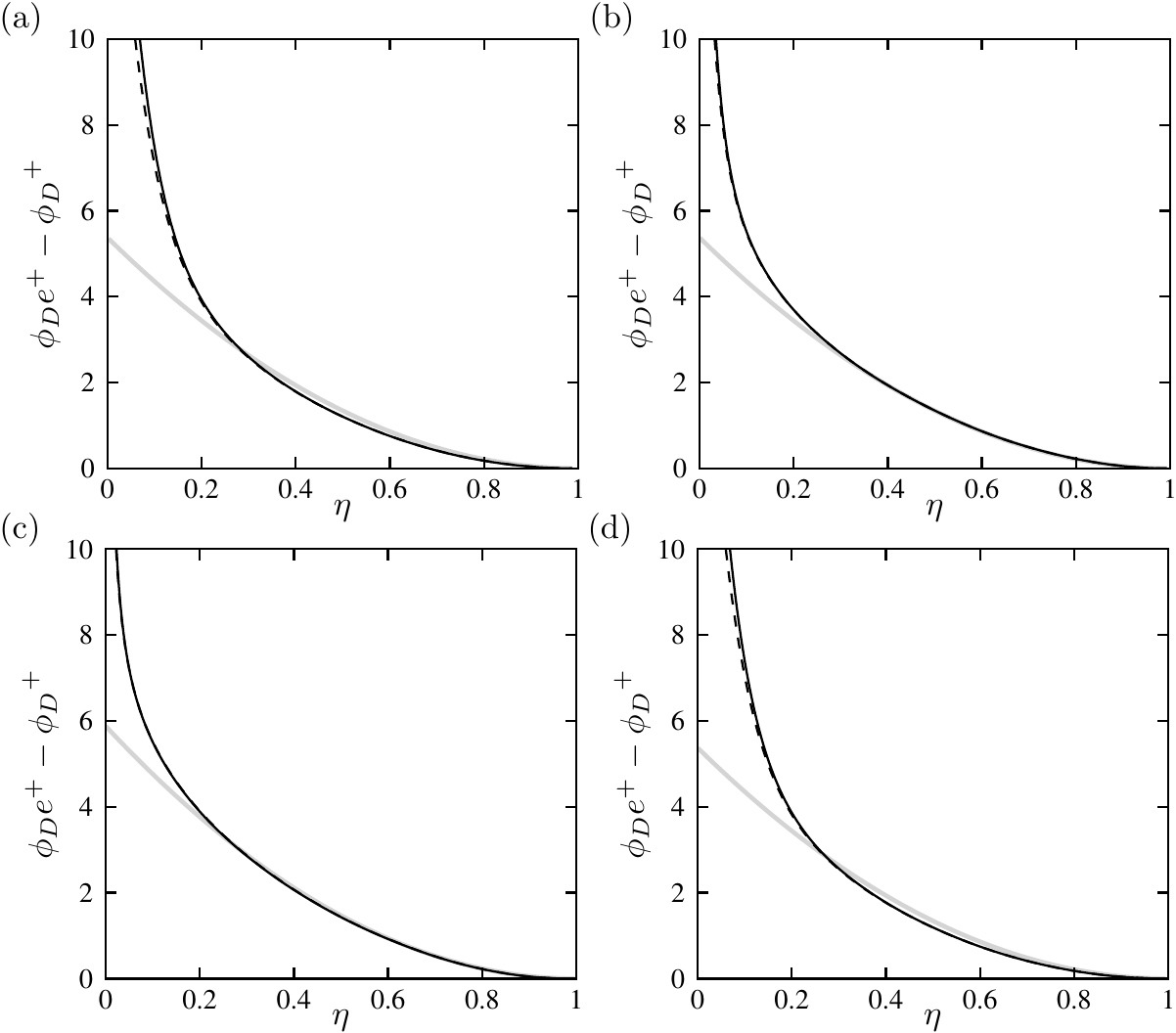}
 \caption{Transformed mean scalar profiles in defect form, as a function
           of $\eta=y/R$ for flow cases
           P15A (a), P15B (b), P15C (c) and P3 (d) at $\Scm=1$ (solid) and $\Scm=0.71$ (dashed).
           The gray lines denote the parabolic velocity profile 
           given in Eqn.~\eqref{eq:outpara_sca}, with $c_\phi=0.0693$.
           The subscript 'e' denotes properties at the pipe centerline.            
  }
 \label{fig:sca_out_pipe}
\end{figure}

As done for the mean velocity we now focus our attention on the mean passive scalar distributions
in the core part of the pipe. \citet{pirozzoli_16} showed that
a universal parabolic profile holds for passive scalar fields in the core part of incompressible
channels. The same reasoning that led us to Eqn.~\eqref{eq:outpara} can then be applied to 
the mean passive scalar balance equation~\eqref{eq:mpassive_pipe}, to obtain
\begin{equation}
 {\phi_{D}^+} -{\phi_{D}^+}_e = -\frac{1}{2c_{\phi}}\left(1-\eta\right)^2,  
 \label{eq:outpara_sca}
\end{equation}
where
\begin{equation}
 {\phi_{D}} = \int_0^{\bar{\phi}}
 {\left(\frac{\bar{\rho}}{\bar{\rho}_w}\right)}^{1/2}
 \mathrm{d}\widetilde{\phi}.
\label{eq:phivd}
\end{equation}
is the mean passive scalar distribution transformed according to van Driest, and the subscript $e$ again
denotes properties at the pipe axis.
Outer defect profiles obtained with van Driest transformation are given
in Fig.~\ref{fig:sca_out_pipe} at $\Scm=1, 0.71$. 
The DNS data are very close to the prediction of Eqn.~\eqref{eq:outpara_sca}
around the pipe centerline,
the range of validity of the parabolic fit extending down to $\eta \approx 0.3$
at sufficiently high Reynolds number.
No clear effects of compressibility and Schmidt number variation 
on the parabolic law constant are observed,
and the latter very nearly coincides with the established incompressible value $c_\phi\approx0.0693$~\citep{pirozzoli_14}.

\subsection{Instantaneous flow field}

\begin{figure}[]
 \centering
 \includegraphics[scale=1.1]{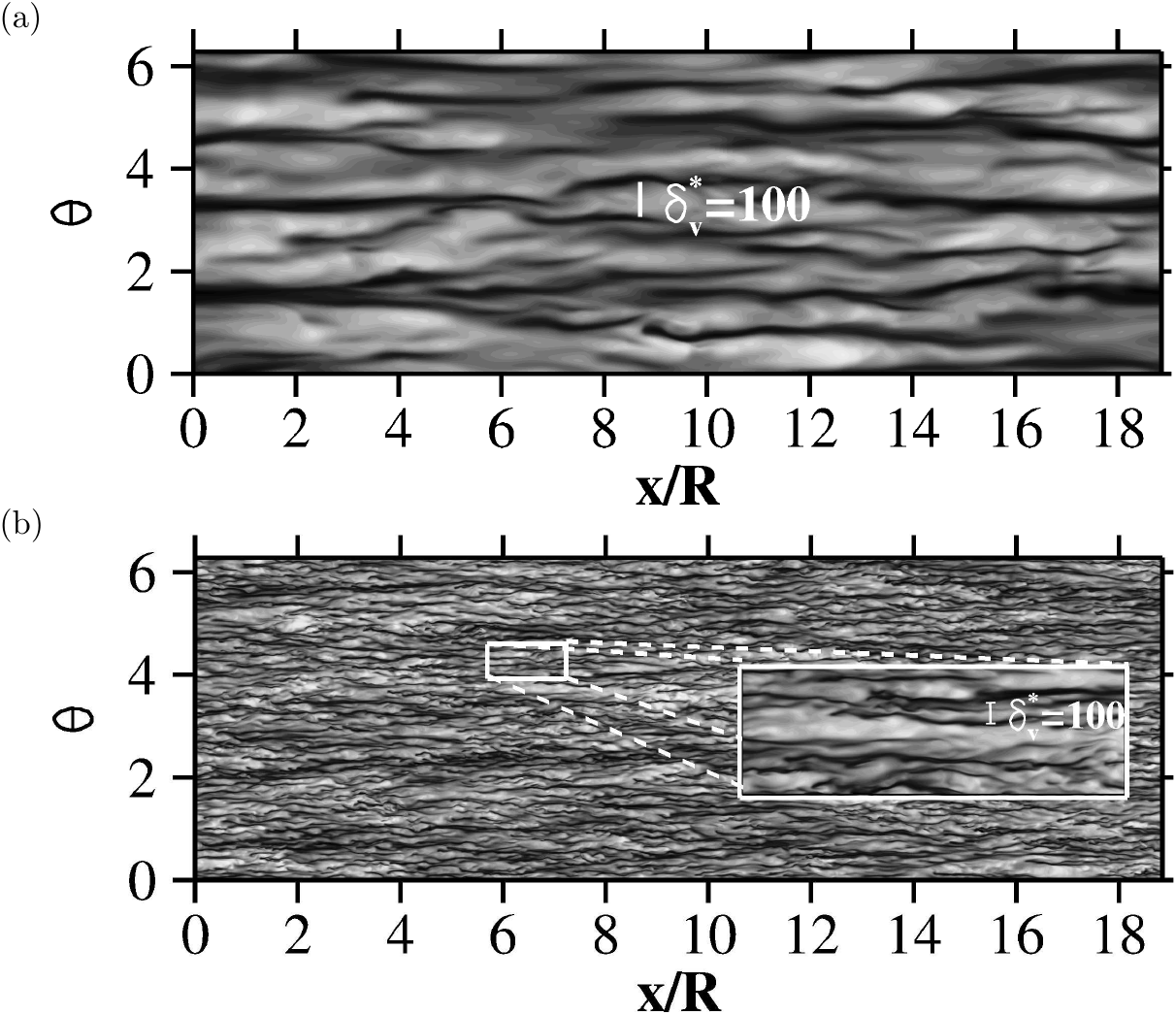}
 \vskip 1em
 \caption{Instantaneous streamwise velocity fluctuations 
          for flow cases P15A (a) and P15C (b) at
          $y_T^+=15$.
          Contours are shown in the range $-4 \le u''/\sqrt{{\tau_{11}}_D}\le 4$,
          from dark to light shades. 
  }
 \label{fig:planexz}
\end{figure}

The general flow organization is scrutinized through visualization of the velocity
and passive scalar fluctuations in the wall-parallel and cross-stream planes.  
Figure~\ref{fig:planexz} shows the instantaneous  streamwise velocity fluctuations
on wall-parallel planes at $y_T^+=15$,
for flow cases P15A and P15C.
The velocity fluctuations in the buffer layer are organized 
in alternating low/high velocity streaky structures, elongated in the streamwise direction.
The figure shows that the azimuthal spacing of the streaks decreases
with the Reynolds number and in particular we find that
the typical spacing is of order $100\delta_v^*$,
the same observed in incompressible flows,
thus further supporting the use of the semi-local scaling~\eqref{eq:SL} 
as effective wall-units in compressible flows.
\begin{figure}[]
 \centering
 \includegraphics[scale=1.1]{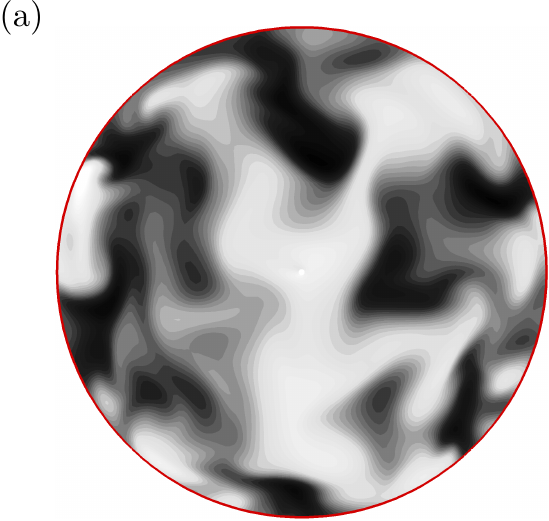}
 \includegraphics[scale=1.1]{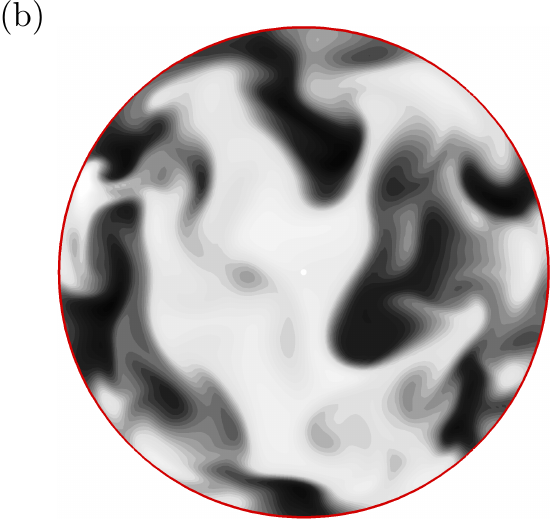}\\
 \includegraphics[scale=1.1]{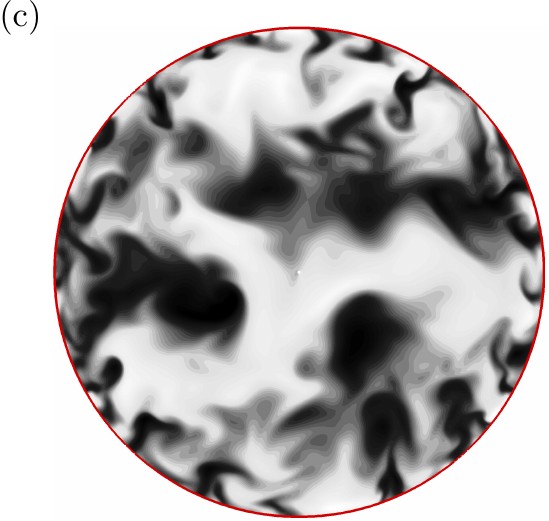}
 \includegraphics[scale=1.1]{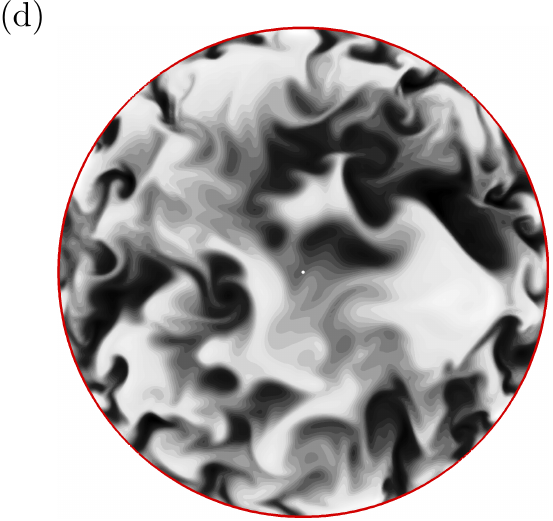}\\
 \includegraphics[scale=1.1]{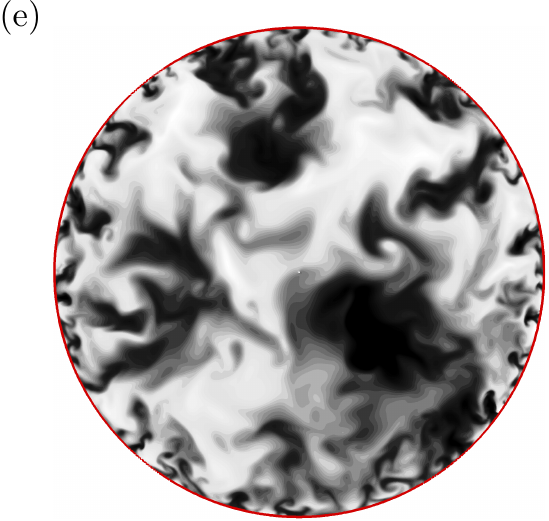}
 \includegraphics[scale=1.1]{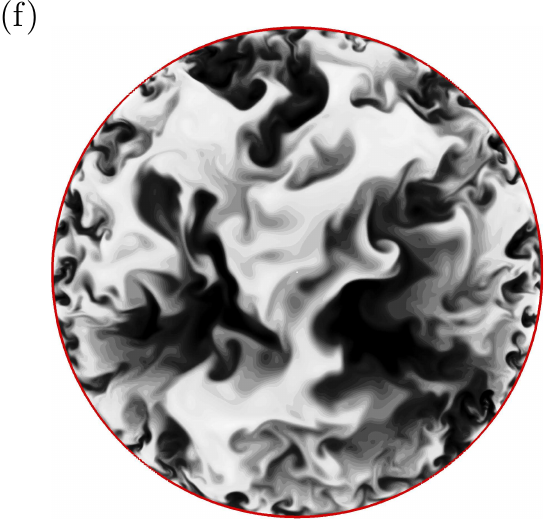}
 \vskip 1em
 \caption{Instantaneous streamwise velocity fluctuations in cross stream plane
          for flow cases P15A (a), P15B (c), P15C (e) 
          and passive scalar fluctuations P15A (b), P15B (d), P15C (f).
          Contours are shown in the range 
          $-4 \le u''/\sqrt{{\tau_{11}}_D}\le 4$ 
          for velocity and 
          $-4 \le \phi''/\sqrt{\left(\widetilde{\phi^{''2}}\right)_D}\le 4$
          for passive scalar at $\Scm=1$, from dark to light shades.
  }
 \label{fig:planeyz}
\end{figure}

A complementary picture of the flow field may be gained by inspecting 
instantaneous flow snapshots in cross-stream planes, shown in Fig.~\ref{fig:planeyz}.
Whereas at low Reynolds number only eddies with $O(R)$ size are found,
scale separation emerges at higher Reynolds number,
with small-scale structures in the near-wall region and $O(R)$ eddies 
in the core region.
As in incompressible wall-bounded flows, the passive scalar field shows substantial 
correlation with the streamwise velocity field, but it 
is characterized by sharper interfaces between regions with positive and negative fluctuations,
owing to the absence of the pressure strain term which tends to smoothen 
the velocity field~\citep{pirozzoli_16}.
The picture thus emerges that the qualitative structure of turbulence in compressible
pipe flow is unaltered by compressibility, and the main features typical
of all incompressible wall-bounded flows are retained.

\subsection{Length scales}

Based on the available DNS data we now turn to evaluating the typical length scales of turbulent eddies 
in the outer part of the wall layer.
In a previous study of compressible channel flow~\citep{modesti_16},
no compressibility effect on the integral length scales was found.
In that study we adapted to compressible flow the idea 
that in the outer layer the typical eddy length scale should depend on the 
friction velocity and the local mean shear~\citep{pirozzoli_12}. 
From dimensional analysis it directly follows, both for velocity and passive scalar, that in compressible flow the relevant length scales for streamwise velocity and passive scalar should be
\begin{eqnarray}
 \ell_{12}^*(y) \sim \left(u_{\tau} R \right)^{1/2}\left(\frac{\mathrm{d}\tilde{u}_D} {\mathrm{d}y}\right)^{-1/2},\quad 
{\ell_{12}}_{\phi}^* (y) \sim \left(\phi_{\tau} R \right)^{1/2} \left( \frac{\mathrm{d}{\phi}_D} {\mathrm{d}y} \right)^{-1/2}. 
\label{eq:l12s}
\end{eqnarray}
where $u_D$ and $\phi_D$ are defined in Eqns.~\eqref{eq:uvd}-\eqref{eq:phivd}, respectively.
It should be noted that in the presence of a logarithmic layer, Eqn.~\eqref{eq:l12s} would yield
the classical scaling $\ell_{12}^*,{\ell_{12}}_{\phi}^* \propto y$ predicted by the attached eddy hypothesis.
In order to evaluate the accuracy of the scaling given in Eqn.~\eqref{eq:l12s},
we consider the spanwise spectral densities of $u$ and $\phi$, defined such that
\begin{equation}
\widetilde{u''^2} = \int_0^{\infty} E_{u}(k_{\theta}) \, \diff k_{\theta}, \quad
\widetilde{\phi''^2} = \int_0^{\infty} E_{\phi}(k_{\theta}) \, \diff k_{\theta}, \quad
\label{eq:PSD}
\end{equation}
where $k_{\theta}$ is the Fourier wavenumber in the azimuthal direction.
To eliminate effects due to variation of the turbulence intensity along the radial direction
we actually consider the normalized spectral densities, defined as
\begin{equation}
\hat{E}_{u}(k_{\theta}) =    {E}_{u}(k_{\theta}) /  \widetilde{u''^2},\quad 
\hat{E}_{\phi}(k_{\theta}) =    {E}_{\phi}(k_{\theta}) /  \widetilde{\phi''^2}. 
\label{eq:PSDnorm}
\end{equation}
For the sake of the analysis, in the following we consider pre-multiplied, normalized energy spectra
as a function of the azimuthal wavelength ($\lambda_{\theta}=2 \pi / k_{\theta}$), 
in semi-log scale. This kind of representation provides hints about the distribution of energy across
the flow scales with the illustrative advantage that equal areas underneath the graphs correspond to equal energies.

\begin{figure}[]
 \centering
 \includegraphics[scale=1.1]{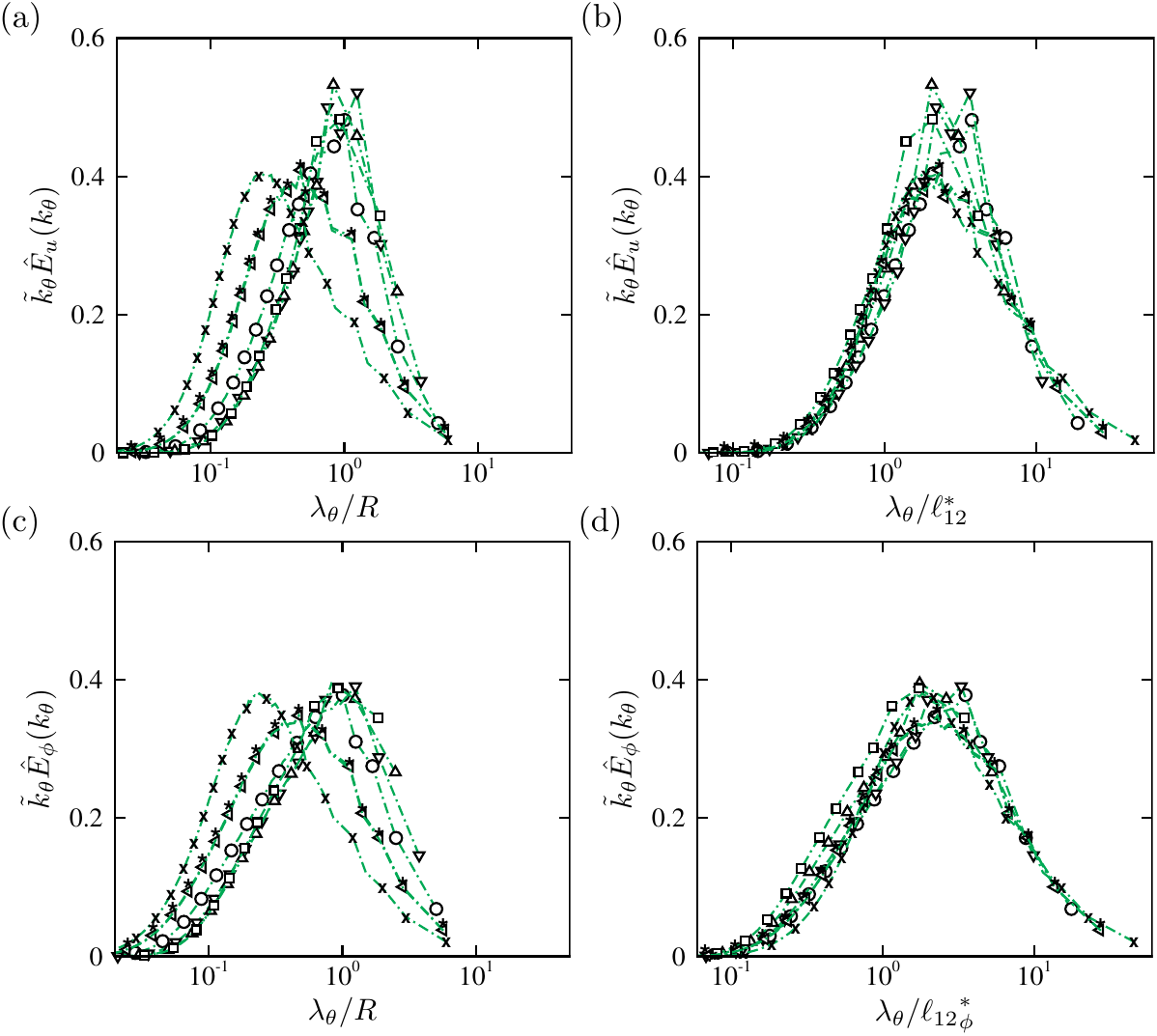}\\
 \vskip 1em
 \caption{
   Flow case P15C: pre-multiplied azimuthal power spectral densities~\eqref{eq:PSDnorm} of $u$ (top row) and $\phi$ (bottom row) 
   as a function of $\lambda_{\theta}/R$ (left column), and of ${\lambda}_{\theta}/\ell_{12}$ (right column). 
   Symbols denote different distances from the wall,
   namely $y^+=50$ (crosses), $y^+=100$ (stars). $\eta=0.1$ (left triangles), $\eta=0.2$ (circles), $\eta=0.4$ (gradients), $\eta=0.6$ (deltas), $\eta=0.7$ (squares).
  }
 \label{fig:specz_pipe}
\end{figure}

Figure~\ref{fig:specz_pipe}(a)-(b) shows pre-multiplied normalized velocity spectra 
at various wall distances across the wall layer, from $y^+=50$ to $\eta=0.7$.
Standard outer scaling as a function of $\lambda_{\theta}/R$ is used in panel (a).
It should be noted that that all spectra have a distinct bump shape,
with energy concentrated at larger and larger scales moving away from the wall.
In particular, in the outer layer, the energy peak corresponds to eddies with azimuthal size ${\lambda}_{\theta}\approx R$, closely resembling the spectral organization observed in incompressible flow~\citep{kim_99,bailey_10}, and consistent with the flow visualizations of Fig.~\ref{fig:planeyz}.
In panel (b) the same spectral densities are shown as a function
of the scaled wavelengths ${\lambda}_{\theta}/\ell_ {12}^*$, ${\ell_{12}}_{\phi}^*$. 
This type of representation does in fact yield significant improvement in the universality 
of the spectral densities, especially as far as the near-wall positions are concerned. The energy 
peak now is found to reside at ${\lambda}_{\theta} = 2-3 \ell_ {12}^*$,
as also found in many incompressible flows~\citep{pirozzoli_16b}.
Based on the established similarity between streamwise velocity and passive scalars,
we expect that the spectra of $u$ and $\phi$ are highly correlated.
Indeed, Fig.~\ref{fig:specz_pipe}(c) shows the same type of organization holds as for 
the streamwise velocity spectra, however the scalar spectra contain more energy at the smallest scales,
owing to the absence of the pressure transfer term in the equation for $\phi$~\citep{pirozzoli_14}, as also observed
in Fig.~\ref{fig:planeyz}.
Similar to what found for the velocity field, we observe that the length scale ${\ell_{12}}_{\phi}^*$
yields improved collapse of the spectra across the wall layer (see panel (d)).

\section{Conclusions}

A DNS database for compressible turbulent flow in a straight circular pipe with isothermal walls
has been developed, achieving Reynolds number significantly higher than previous studies,
thus allowing to observe typical features of developed wall turbulence as a genuine logarithmic layer
in the mean velocity distribution. The flow solver relies on a baseline central discretization
of the convective derivatives expanded to quasi-skew-symmetric form, in such a way that
discrete preservation of the total kinetic energy from convection is guaranteed.
Efficient time stepping is guaranteed through semi-implicit treatment of acoustic
waves in the convective terms, and local azimuthal coarsening near the pipe symmetry axis.
Preliminary validation studies support good predictive capabilities of the solver.

Analysis of the velocity field generally supports findings of previous studies carried out in 
planar channel flow~\citep{modesti_16}. In particular, we find that Huang's transformation
for the wall-normal coordinate coupled with standard density scaling yields near universality of the
Reynolds stress components, the only exception being the peak streamwise velocity variance.
Partial failure of Morkovin's hypothesis in this case is traditionally recognized as one
of the few genuine compressibility effects, and sometimes traced to modifications of the
pressure-strain term in the streamwise momentum equation~\citep{foysi_04}.
However, we find that the same effect is also present in the passive scalar variance
distributions also upon density scaling (see Fig.~\ref{fig:scal_fluc_pipe}), in which the
pressure term is no present. Hence, reasons for this (limited) lack of similarity
may be even more subtle than previously thought.
Van Driest transformation is found to be inaccurate in yielding universality of the
mean velocity distributions, whereas the Trettel-Larsson transformation
better accounts for mean density and viscosity variations taking place within the buffer layer.
Excellent collapse with incompressible pipe data and compressible channel data is
then achieved, and a sizeable canonical logarithmic layer is found to form at $\Rey_{\tau} \approx 1000$.
The core part of the pipe is found to have a relatively simple structure, whereby the
van Driest transformed velocity follows a universal parabolic law in a
wide region, and it is controlled by a single universal constant, here
found to be identical to that of incompressible pipe flow.
As in other compressible wall-bounded flows, the mean temperature distribution
is found to quadratically depend on the mean streamwise velocity,
and we find that the generalized Reynolds analogy of \citet{zhang_14} yields
very good prediction of its variation.
Together with TL transformation, use of the temperature/velocity relation yields
a closed system of equations, which in principle lends itself to closed formulas
for the prediction of the friction and heat transfer coefficients, which might
be the subject of further investigations.

The analysis of passive scalars in turbulent flow is interesting in its own sake,
and because it allows to establish similarities/differences with respect
to the streamwise velocity and temperature fields, which are governed by formally
similar equations. Similarity with the velocity field are in fact strong,
and we show that the TL transformation can be adapted to predict the mean incompressible
passive scalar distributions. Temperature has on the other hand a very
different behavior than a passive scalar, owing to strong spatial inhomogeneity of the
viscous heating term.
Finally, we have analyzed velocity and passive scalar spectra with the goal of establishing
a parametrization for the growth of the typical eddy size along the wall-normal direction.
In this respect we find that good universality of the spectral distributions is achieved when
wavelengths are scaled with a length scale constructed with the local friction
velocity defined in Eqn.~\eqref{eq:SL}, and the local mean shear based on the van Driest
transformed mean velocity profile. The same conclusion also applies to passive scalars.\\
{\bf Acknowledgements}\\
We acknowledge that most of the results reported in this paper have been achieved using the PRACE Research Infrastructure resource MARCONI based at CINECA, Casalecchio di Reno, Italy.

\addcontentsline{toc}{chapter}{Bibliography}
\bibliographystyle{plainnat}
\bibliography{references} 

\begin{thebibliography}{46}
\providecommand{\natexlab}[1]{#1}
\providecommand{\url}[1]{\texttt{#1}}
\expandafter\ifx\csname urlstyle\endcsname\relax
  \providecommand{\doi}[1]{doi: #1}\else
  \providecommand{\doi}{doi: \begingroup \urlstyle{rm}\Url}\fi

\bibitem[Ahn et~al.(2015)Ahn, Lee, Lee, Kang, and Sung]{ahn_15}
J.~Ahn, J.H. Lee, J.~Lee, J.H. Kang, and H.J. Sung.
\newblock Direct numerical simulation of a $30{R}$ long turbulent pipe flow at
  ${R}e_\tau= 3008$.
\newblock \emph{Phys. Fluids (1994-present)}, 27\penalty0 (6):\penalty0 065110,
  2015.

\bibitem[Alamo and Jim{\'e}nez(2009)]{del_09}
J.C.~Del Alamo and J.~Jim{\'e}nez.
\newblock Estimation of turbulent convection velocities and corrections to
  {T}aylor's approximation.
\newblock \emph{J. Fluid Mech.}, 640:\penalty0 5--26, 2009.

\bibitem[Bailey and Smits(2010)]{bailey_10}
S.C.C. Bailey and A.J. Smits.
\newblock Experimental investigation of the structure of large-and
  very-large-scale motions in turbulent pipe flow.
\newblock \emph{J.~Fluid~Mech.}, 651:\penalty0 339--356, 2010.

\bibitem[Bogey et~al.(2011)Bogey, Cacqueray, and Bailly]{bogey_11_jcp}
C.~Bogey, N.~De Cacqueray, and C.~Bailly.
\newblock Finite differences for coarse azimuthal discretization and for
  reduction of effective resolution near origin of cylindrical flow equations.
\newblock \emph{J. Comput. Phys.}, 230\penalty0 (4):\penalty0 1134--1146, 2011.

\bibitem[Chin et~al.(2014)Chin, Monty, and Ooi]{chin_14}
C.~Chin, J.P. Monty, and A.~Ooi.
\newblock Reynolds number effects in {DNS} of pipe flow and comparison with
  channels and boundary layers.
\newblock \emph{Int. J. Heat Fluid Flow}, 45:\penalty0 33--40, 2014.

\bibitem[Chin et~al.(2015)Chin, Ng, Blackburn, Monty, and Ooi]{chin_15}
C.~Chin, H.C.H Ng, H.M. Blackburn, J.P. Monty, and A.~Ooi.
\newblock Turbulent pipe flow at ${R}e_\tau\approx 1000$: A comparison of
  wall-resolved large-eddy simulation, direct numerical simulation and hot-wire
  experiment.
\newblock \emph{Comp. Fluids}, 122:\penalty0 26--33, 2015.

\bibitem[Coleman et~al.(1995)Coleman, Kim, and Moser]{coleman_95}
G.N. Coleman, J.~Kim, and R.D. Moser.
\newblock A numerical study of turbulent supersonic isothermal-wall channel
  flow.
\newblock \emph{J.\ Fluid\ Mech.}, 305:\penalty0 159--183, 1995.

\bibitem[Duan et~al.(2010)Duan, Beekman, and Martin]{duan_10}
L.~Duan, I.~Beekman, and M.P. Martin.
\newblock Direct numerical simulation of hypersonic turbulent boundary layers.
  {P}art 2. {E}ffect of wall temperature.
\newblock \emph{J.\ Fluid\ Mech.}, 655:\penalty0 419--445, 2010.

\bibitem[Eggels et~al.(1994)Eggels, Unger, Weiss, Westerweel, Adrian,
  Friedrich, and Nieuwstadt]{eggels_94}
J.G.M. Eggels, F.~Unger, M.H. Weiss, J.~Westerweel, R.J. Adrian, R.~Friedrich,
  and F.T.M. Nieuwstadt.
\newblock Fully developed turbulent pipe flow: a comparison between direct
  numerical simulation and experiment.
\newblock \emph{J. Fluid Mech.}, 268:\penalty0 175--210, 1994.

\bibitem[Foysi and Friedrich(2005)]{foysi_05}
H.~Foysi and R.~Friedrich.
\newblock Passive scalar transport in turbulent supersonic channel flow.
\newblock In \emph{Progr. Turbulence}, pages 223--227. Springer, 2005.

\bibitem[Foysi et~al.(2004)Foysi, Sarkar, and Friedrich]{foysi_04}
H.~Foysi, S.~Sarkar, and R.~Friedrich.
\newblock Compressibility effects and turbulence scalings in supersonic channel
  flow.
\newblock \emph{J.\ Fluid\ Mech.}, 509:\penalty0 207--216, 2004.

\bibitem[Furuichi et~al.(2015)Furuichi, Terao, Wada, and Tsuji]{furuichi_15}
N.~Furuichi, Y.~Terao, Y.~Wada, and Y.~Tsuji.
\newblock Friction factor and mean velocity profile for pipe flow at high
  {R}eynolds numbers.
\newblock \emph{Phys. Fluids (1994-present)}, 27\penalty0 (9):\penalty0 095108,
  2015.

\bibitem[Ghosh et~al.(2006)Ghosh, Sesterhenn, and Friedrich]{ghosh_06}
S.~Ghosh, J.~Sesterhenn, and R.~Friedrich.
\newblock {DNS} and {LES} of compressible turbulent pipe flow with isothermal
  wall.
\newblock In \emph{Direct and Large-Eddy Simulation VI}, pages 721--728.
  Springer, 2006.

\bibitem[Ghosh et~al.(2010)Ghosh, Foysi, and Friedrich]{ghosh_10}
S.~Ghosh, H.~Foysi, and R.~Friedrich.
\newblock Compressible turbulent channel and pipe flow: similarities and
  differences.
\newblock \emph{J. Fluid Mech.}, 648:\penalty0 155--181, 2010.

\bibitem[Huang et~al.(1995)Huang, Coleman, and Bradshaw]{huang_95}
P.G. Huang, G.N. Coleman, and P.~Bradshaw.
\newblock Compressible turbulent channel flows: {DNS} results and modeling.
\newblock \emph{J.\ Fluid\ Mech.}, 305:\penalty0 185--218, 1995.

\bibitem[Kader(1981)]{kader_81}
B.A. Kader.
\newblock Temperature and concentration profiles in fully turbulent boundary
  layers.
\newblock \emph{Int. J. Heat Mass Transf.}, 24\penalty0 (9):\penalty0
  1541--1544, 1981.

\bibitem[Kim and Adrian(1999)]{kim_99}
K.C. Kim and R.J. Adrian.
\newblock Very large-scale motion in the outer layer.
\newblock \emph{Phys. Fluids}, 11\penalty0 (2):\penalty0 417--422, 1999.

\bibitem[Kjellstr{\"o}m and Hedberg(1968)]{kjellstrom_68}
B.~Kjellstr{\"o}m and S.~Hedberg.
\newblock Calibration experiments with a {DISA} hot-wire anemometer.
\newblock Technical report, AB Atomenergi, 1968.

\bibitem[Klein et~al.(2003)Klein, Sadiki, and Janicka]{klein_03}
M.~Klein, A.~Sadiki, and J.~Janicka.
\newblock A digital filter based generation of inflow data for spatially
  developing direct numerical or large eddy simulations.
\newblock \emph{J. Comput. Phys.}, 186\penalty0 (2):\penalty0 652--665, 2003.

\bibitem[Lee et~al.(2015)Lee, Ahn, and Sung]{lee_15}
J.~Lee, J.~Ahn, and H.J. Sung.
\newblock Comparison of large-and very-large-scale motions in turbulent pipe
  and channel flows.
\newblock \emph{Phys. Fluids (1994-present)}, 27\penalty0 (2):\penalty0 025101,
  2015.

\bibitem[Mckeon et~al.(2004)Mckeon, Li, Jiang, Morrison, and Smits]{mckeon_04}
B.J. Mckeon, J.~Li, W.~Jiang, J.F. Morrison, and A.J. Smits.
\newblock Further observations on the mean velocity distribution in fully
  developed pipe flow.
\newblock \emph{J. Fluid Mech.}, 501:\penalty0 135--147, 2004.

\bibitem[Modesti and Pirozzoli(2016)]{modesti_16}
D.~Modesti and S.~Pirozzoli.
\newblock Reynolds and {M}ach number effects in compressible turbulent channel
  flow.
\newblock \emph{Int. J. Heat Fluid Flow}, 59:\penalty0 33--49, 2016.

\bibitem[Modesti et~al.()Modesti, Pirozzoli, and Grasso]{modesti_18}
D.~Modesti, S.~Pirozzoli, and F.~Grasso.
\newblock Direct numerical simulation of developed compressible flow in square
  ducts.
\newblock \emph{arXiv preprint arXiv:1808.00282}.
\newblock Submitted to Int. J. Heat Fluid Flow.

\bibitem[Mohseni and Colonius(2000)]{mohseni_00}
K.~Mohseni and T.~Colonius.
\newblock Numerical treatment of polar coordinate singularities.
\newblock \emph{J. Comput. Phys.}, 157\penalty0 (2):\penalty0 787--795, 2000.

\bibitem[Morkovin(1962)]{morkovin_62}
M.V. Morkovin.
\newblock Effects of compressibility on turbulent flows.
\newblock In \emph{M{\'e}canique de la Turbulence}, pages 367--380. A. {F}avre,
  1962.

\bibitem[Orlandi and Fatica(1997)]{orlandi_97}
P.~Orlandi and M.~Fatica.
\newblock Direct simulations of turbulent flow in a pipe rotating about its
  axis.
\newblock \emph{J. Fluid Mech.}, 343:\penalty0 43--72, 1997.

\bibitem[Patel and Head(1969)]{patel_69}
V.C. Patel and M.R. Head.
\newblock Some observations on skin friction and velocity profiles in fully
  developed pipe and channel flows.
\newblock \emph{J. Fluid Mech.}, 38\penalty0 (1):\penalty0 181--201, 1969.

\bibitem[Pirozzoli(2010)]{pirozzoli_10}
S.~Pirozzoli.
\newblock Generalized conservative approximations of split convective
  derivative operators.
\newblock \emph{J. Comput. Phys.}, 229\penalty0 (19):\penalty0 7180--7190,
  2010.

\bibitem[Pirozzoli(2011)]{pirozzoli_11}
S.~Pirozzoli.
\newblock Stabilized non-dissipative approximations of {E}uler equations in
  generalized curvilinear coordinates.
\newblock \emph{J. Comput. Phys.}, 230\penalty0 (8):\penalty0 2997--3014, 2011.

\bibitem[Pirozzoli(2012)]{pirozzoli_12}
S.~Pirozzoli.
\newblock On the size of the energy-containing eddies in the outer turbulent
  wall layer.
\newblock \emph{J.\ Fluid\ Mech.}, 702:\penalty0 521--532, 2012.

\bibitem[Pirozzoli(2014)]{pirozzoli_14}
S.~Pirozzoli.
\newblock Revisiting the mixing-length hypothesis in the outer part of
  turbulent wall layers: mean flow and wall friction.
\newblock \emph{J.\ Fluid\ Mech.}, 745:\penalty0 378--397, 2014.

\bibitem[Pirozzoli(2016)]{pirozzoli_16b}
S.~Pirozzoli.
\newblock On the {S}ize of the {E}ddies in the {O}uter {T}urbulent {W}all
  {L}ayer: {E}vidence from {V}elocity {S}pectra.
\newblock In \emph{Progress in Wall Turbulence 2}, pages 3--15. Springer, 2016.

\bibitem[Pirozzoli and Bernardini(2013)]{pirozzoli_13}
S.~Pirozzoli and M.~Bernardini.
\newblock Probing high-{R}eynolds-number effects in numerical boundary layers.
\newblock \emph{Phys.~Fluids}, 25\penalty0 (2):\penalty0 021704, 2013.

\bibitem[Pirozzoli et~al.(2016)Pirozzoli, Bernardini, and
  Orlandi]{pirozzoli_16}
S.~Pirozzoli, M.~Bernardini, and P.~Orlandi.
\newblock Passive scalars in turbulent channel flow at high {R}eynolds number.
\newblock \emph{J. Fluid Mech.}, 788:\penalty0 614--639, 2016.

\bibitem[Sandberg et~al.(2012)Sandberg, Sandham, and Suponitsky]{sandberg_12}
R.D. Sandberg, N.D. Sandham, and V.~Suponitsky.
\newblock {DNS} of compressible pipe flow exiting into a coflow.
\newblock \emph{Int. J. Heat Fluid Flow}, 35:\penalty0 33--44, 2012.

\bibitem[Smits and Dussauge(1996)]{smits_96}
A.J. Smits and J.-P. Dussauge.
\newblock \emph{Turbulent shear layers in supersonic flow, 2nd Edn.}
\newblock American Istitute of Physics, 1996.

\bibitem[Spina et~al.(1994)Spina, Smits, and Robinson]{spina_94}
E.F. Spina, A.J. Smits, and S.K. Robinson.
\newblock The physics of supersonic turbulent boundary layers.
\newblock \emph{Annu.\ Rev.\ Fluid\ Mech.}, 26:\penalty0 287--319, 1994.

\bibitem[Trettel and Larsson(2016)]{trettel_16}
A.~Trettel and J.~Larsson.
\newblock Mean velocity scaling for compressible wall turbulence with heat
  transfer.
\newblock \emph{Phys. Fluids (1994-present)}, 28\penalty0 (2):\penalty0 026102,
  2016.

\bibitem[van Driest(1951)]{vandriest_51}
E.R. van Driest.
\newblock Turbulent boundary layer in compressible fluids.
\newblock \emph{J.\ Aero.\ Sci.}, 18:\penalty0 145--160, 1951.

\bibitem[Verzicco and Orlandi(1996)]{verzicco_96}
R.~Verzicco and P.~Orlandi.
\newblock A finite-difference scheme for three-dimensional incompressible flows
  in cylindrical coordinates.
\newblock \emph{J. Comput. Phys.}, 123\penalty0 (2):\penalty0 402--414, 1996.

\bibitem[Walz(1959)]{walz_59}
A.~Walz.
\newblock Compressible turbulent boundary layers with heat transfer and
  pressure gradient in flow direction.
\newblock \emph{J.\ Res.\ Natl.\ Bur.\ Stand.}, 63, 1959.

\bibitem[Williams et~al.(2018)Williams, Sahoo, Baumgartner, and
  Smits]{williams_18}
O.J.H. Williams, D.~Sahoo, M.L. Baumgartner, and A.J. Smits.
\newblock Experiments on the structure and scaling of hypersonic turbulent
  boundary layers.
\newblock \emph{J. Fluid Mech.}, 834:\penalty0 237--270, 2018.

\bibitem[Wu and Moin(2008)]{wu_08}
X.~Wu and P.~Moin.
\newblock A direct numerical simulation study on the mean velocity
  characteristics in turbulent pipe flow.
\newblock \emph{J. Fluid Mech.}, 608:\penalty0 81--112, 2008.

\bibitem[Wu et~al.(2012)Wu, Baltzer, and Adrian]{wu_12}
X.~Wu, J.R. Baltzer, and R.J. Adrian.
\newblock Direct numerical simulation of a $30{R}$ long turbulent pipe flow at
  ${R}e_\tau= 685$: large-and very large-scale motions.
\newblock \emph{J. Fluid Mech.}, 698:\penalty0 235--281, 2012.

\bibitem[Zagarola and Smits(1998)]{zagarola_98}
M.V. Zagarola and A.J. Smits.
\newblock Mean-flow scaling of turbulent pipe flow.
\newblock \emph{J. Fluid Mech.}, 373:\penalty0 33--79, 1998.

\bibitem[Zhang et~al.(2014)Zhang, Bi, Hussain, and She]{zhang_14}
Y.S. Zhang, W.T. Bi, F.~Hussain, and Z.S. She.
\newblock A generalized {R}eynolds analogy for compressible wall-bounded
  turbulent flows.
\newblock \emph{J.\ Fluid\ Mech.}, 739:\penalty0 392--420, 2014.

\end{thebibliography}
\end{document}